\documentclass{jpp}
\usepackage[latin1]{inputenc}
\usepackage{amsmath}
\usepackage{amsfonts}
\usepackage{amssymb}
\usepackage{mathtools, cuted}
\usepackage{graphicx}
\usepackage{natbib}
\usepackage{placeins}

\begin{document}
\title{Parallel expansion of a fuel pellet plasmoid}
\author{Alistair M. Arnold\aff{1}\corresp{\email{alistair.arnold@ipp.mpg.de}},
Pavel Aleynikov\aff{1}, Boris N. Breizman\aff{2}}
\affiliation{\aff{1}Max-Planck-Institut f\"{u}r Plasmaphysik, Greifswald, Germany
\aff{2}Institute for Fusion Studies, University of Texas at Austin, Austin, USA}
\maketitle

\section*{Abstract}
The problem of the expansion and assimilation of a cryogenic fuel pellet injected into a hot plasma is considered. Due to the transparency of the plasmoid to ambient particles, it is found that electrons reach a `quasi-equilibrium' (QE) which is characterised by a steady-state on the fastest collisional timescale. The simplified electron kinetic equation of the quasi-equilibrium state is solved. Taking a velocity moment of the electron kinetic equation permits a fluid closure, yielding an evolution equation for the parameters describing the QE distribution function. In contrast to the Braginskii equations, the closure does not require that electrons have a short mean free path compared to the size of density perturbations and permits an anisotropic and highly non-Maxwellian distribution function. Since the QE electron distribution function accounts for both trapped and passing electrons, the self-consistent electric potential that causes the expansion can be properly described, in contrast to earlier models of pellet plasmoid expansion with an unbounded potential. The plasmoid expansion is simulated using both a Vlasov model and a cold fluid model for the ions. During the expansion plasmoid ions and electrons obtain a nearly equal amount of energy; as hot ambient electrons provide this energy in the form of collisional heating of plasmoid electrons, the expansion of a pellet plasmoid is expected to be a potent mechanism for the transfer of energy from electrons to ions on a timescale shorter than that of ion-electron thermalisation.

\section{Introduction}
During a recent experimental campaign of the W7-X stellarator, fuel pellet injection was found to be associated with a large transfer of energy from electrons to ions \citep{Baldzuhn2019,Baldzuhn2020,Bozhenkov2020}. Such an energy transfer is generally desirable, since it acts to bring the ion temperature up to the electron temperature (the ion temperature being lower than the electron temperature during normal operation), and a higher ion temperature results in a larger fusion cross-section. Subsequent investigation of the dynamics of the injection and assimilation of fuel pellets have suggested that a possible mechanism for the energy transfer is the rapid ambipolar expansion of the ionised pellet material -- the pellet plasmoid -- along magnetic field lines \citep{Aleynikov2019,Runov2021,Arnold2021}. The aim of this paper is to resolve the inconsistencies present in these models and provide a rigorous model of the parallel plasmoid expansion. What follows is a brief recapitulation of the processes by which the pellet plasmoid is formed, the reason for its parallel expansion and concomitant electron-ion energy transfer, and a summary of the approaches and pitfalls of earlier models. An outline of a new approach which does not suffer from these pitfalls is provided before being realised mathematically.

%However, These investigations assumed a near-Maxwellian electron distribution function. In this paper a more rigorous approach to electron kinetics is taken, resulting in a description that does not suffer from the pitfalls of earlier investigations, the most severe of which being the unbounded electric potential far from the pellet material. Additionally we consider both collisionless and highly collisional models for the ions.

When a fuel pellet is injected into an MCF (Magnetic Confinement Fusion) device, the incoming energy flux from the ambient plasma ablates the surface of the pellet and produces a gas cloud \citep{Parks1977}. The pellet and gas are composed of electrically neutral molecules, but plasma is continuously generated within the gas cloud by the collisions of the high-energy ions and electrons composing the multi $\mathrm{keV}$ ambient plasma with gas molecules. Subsequently, the pellet and gas cloud continue to cross magnetic field lines at the speed at which they were injected, but some of the newly-ionised plasma is left behind; that which was not collisionally `dragged' along with the moving gas cloud. This is because the plasma constituents are charged particles and follow Larmor orbits that `pin' the particles to the field line.  The result is that a \textit{plasmoid}, a localised excess density of plasma, is deposited on field lines that intersected the gas cloud as it traversed the device.

\begin{figure}
    \centering
    \includegraphics[width=0.8\textwidth]{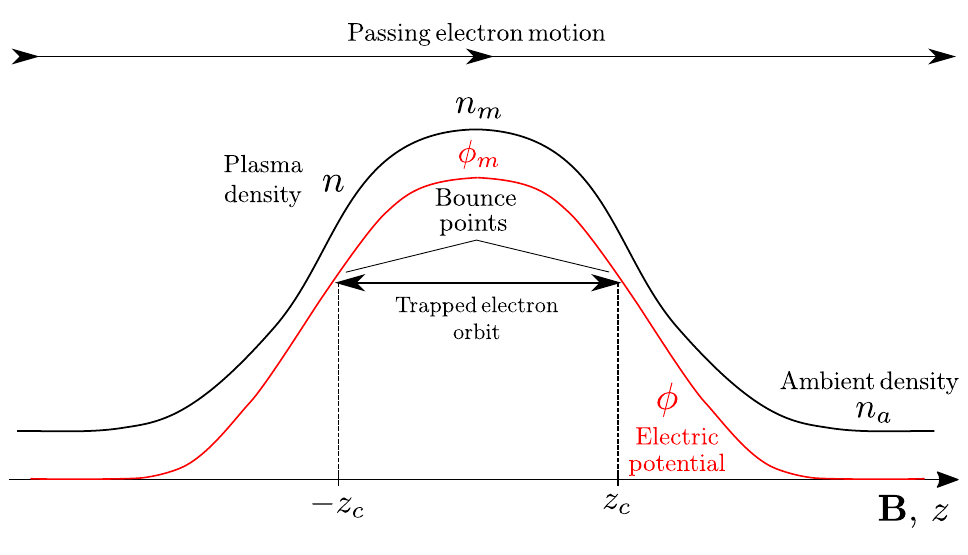}
    \caption{Schematic of the electric potential induced by the presence of the plasmoid  \citep{Arnold2023}. Example trapped (with turning points $\pm z_c$) and passing electron trajectories are included. The profiles are assumed to be even and monotonically decreasing in $z$ with the electron density and potential reaching their maxima $n_m$ and $\phi_m$ at $z = 0$.}
    \label{fig:well_schematic}
\end{figure}

Since the plasmoid is a localised density perturbation and electrons have a much higher thermal velocity than ions, the electric potential required to maintain quasineutrality acts to trap electrons inside the plasmoid and accelerate ions away from the plasmoid. Figure \ref{fig:well_schematic} shows a schematic of the plasmoid and electric potential. Since the potential acts to trap electrons, we will use the names `well' and `potential' interchangeably. With regards to pellet plasmoids the density is such that the the electric potential drives parallel dynamics much more quickly than transverse dynamics occur, the latter being due to drifts. Therefore, as with previous investigations, we consider only the parallel expansion of the pellet plasmoid on one given field line.

We stress that not all of the plasma produced from the pellet ablatant fits the description of the previous paragraph. Naturally, plasma that is `dragged along' with the gas cloud exhibits quite different dynamics. However, for `fast' pellet injection devices proportionally more of the plasmoid dynamics occur on field lines where the gas cloud and pellet have departed \citep{Arnold2021}. The injection devices fitting the criteria of being `fast' are becoming the norm in MCF experiments, so the conclusions drawn from the parallel plasmoid expansion in the absence of gas can be expected to reasonably well apply to pellet injection in future MCF devices.

The dynamics of any plasmoid immersed in an ambient plasma depend greatly upon the plasma and plasmoid parameters, such as their relative temperatures, densities, the plasmoid size, and so on. Naturally, it is difficult to describe plasmoid dynamics with a too wide-ranging choice of parameters, so our attention must be restricted to plasmoids broadly corresponding to a those produced by pellet injection in a state-of-the-art MCF device. We take as a reference point the W7-X stellarator, since the success of its pellet injection campaign provides motivation for studying pellet plasmoids. Further, the temperatures and densities in the core of W7-X are generally comparable to other high-performance MCF devices.

For the purpose of untangling the different phenomena involved in plasmoid expansion it is helpful to provide concrete plasma parameters We consider an ambient plasma of electron density $n_a = 5\times10^{19}\,\mathrm{m}^{-3}$ at a temperature $T_a = 5\,\mathrm{keV}$. A typical line-integrated density along the field line of a fuel pellet plasmoid in W7-X is $N_p = 10^{22}\,\mathrm{m}^{-2}$ \citep{Arnold2021}. 

The fuel pellets in W7-X contain approximately $10^{20}$ electrons and penetrate roughly $0.1\,\mathrm{m}$ into the plasma, resulting in the average line-density in the radial direction of $10^{21}\,\mathrm{m}^{-1}$ \citep{Baldzuhn2019}. In W7-X the flux surface located at minor radius $r = 0.3\,\mathrm{m}$, given the major radius $R = 5\,\mathrm{m}$, has a flux-surface integrated density of $3\times10^{21}\,\mathrm{m}^{-1}$ if the density of this flux surface is $5\times10^{19}\,\mathrm{m}^{-3}$. Hence for such a flux surface the temperature is not strongly quenched after assimilation. For high-performance scenarios in W7-X the quenching is even weaker due to the higher plasma densities. Therefore, unlike killer pellets, which quench the temperature completely, on large flux surfaces fuelling pellets only slightly affect the temperature. We therefore neglect any change in $T_a$ during the plasmoid expansion.

We consider irrational flux surfaces where individual field lines have a connection length $L_F$ that is, in principle, infinite. However, given that the plasmoid has a transverse size $r_I$, the connection length of the flux tube containing the plasmoid is in practice $(2\pi R)(2\pi r)/r_I$, since this is the length after which the flux tube of diameter $r_I$ self-intersects. For W7-X pellets $r_I \approx 0.1\,\mathrm{m}$ shortly after injection \citep{Baldzuhn2019,Arnold2021}, giving a maximum connection length of $600\,\mathrm{m}$. Since the plasmoid is expected to reach this size after its density has dropped to practically the value of the ambient plasma for $N_p = 10^{22}\,\mathrm{m}^{-2}$ \citep{Aleynikov2019}, we formally take the connection length to be infinite. \cite{Arnold2023} in contrast treated the electron kinetics in a high-Z plasmoid on a field line of finite connection length, accounting for the quenching of the ambient plasma.

Since plasmoid electrons are `born' at energies comparable to the ionisation energy, of order tens of $\mathrm{eV}$, but are immersed in an ambient plasma with a temperature on the order of several $\mathrm{keV}$, the electron distribution function as a whole will consist of a cold, dense core of plasmoid electrons and a hot, sparse tail of ambient electrons. The distribution function is only close to a Maxwellian after the plasmoid electrons have been sufficiently heated by the ambient electrons, which happens after the plasmoid has significantly expanded with the plasma parameters we use here. The primary concern with previous models of the expansion is that they treated only the cold plasmoid electrons, assuming that they have a near-Maxwellian distribution function, but did not treat ambient electrons. These electrons were simply assumed to be of a constant density, hence providing collisional heating to the plasmoid electrons. Further, ambient ions were not considered at all in the fluid model for the ions. The consequence of this approach is that the electric potential decreases without bound as the plasmoid density vanishes. Clearly, one cannot use this approach as a basis for treating both trapped and passing electrons. The fact that the electric potential was unphysical also called into question the result of the electron-ion energy transfer and other aspects of the expansion.

\begin{figure}
    \centering
    \includegraphics[width=0.8\textwidth]{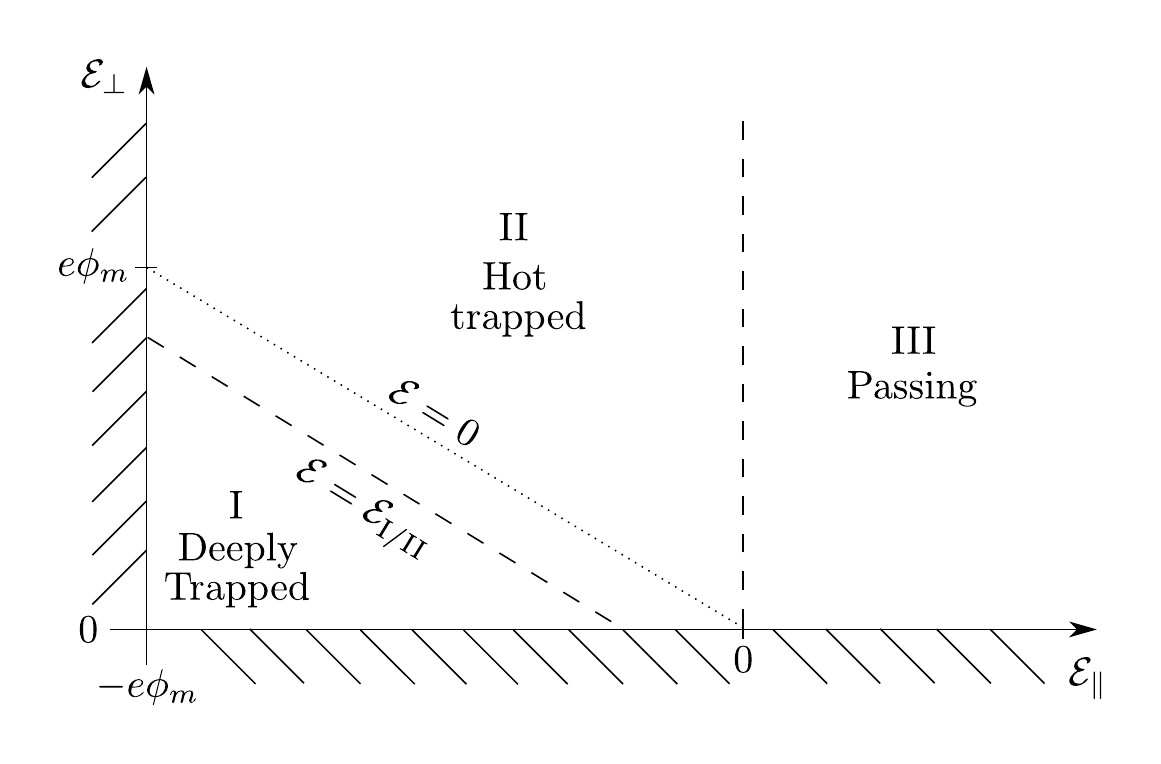}
    \caption{Schematic of the phase-space domain of the electron kinetic problem at $z = 0$. This is also the domain for the bounce-averaged kinetic problems. The dotted line indicates $\mathcal{E} = \mathcal{E}_\parallel + \mathcal{E}_\perp = 0$. The diagonal dashed line indicates $\mathcal{E} = \mathcal{E}_\mathrm{I/II}$, which separates regions I and II. The vertical dashed line indicates $\mathcal{E}_\parallel = 0$, the trapped-passing separatrix.}
    \label{fig:domain-schematic}
\end{figure}

A more sophisticated approach to electrons is required to resolve these issues. We will consider electron kinetics in the variables of parallel energy $\mathcal{E}_\parallel = m_e v_\parallel^2/2 -e \phi(z)$, where $v_\parallel$ is the velocity parallel to the field line, perpendicular energy $\mathcal{E}_\perp = m_e v_\perp^2/2$, where $v_\perp$ is the speed perpendicular to the field lines, $z$, the position along the field line, and $t$, time. In anticipation of the form of the distribution function for different energies we split the phase space into regions I, II, and III, respectively corresponding to the deeply trapped electrons, hot trapped electrons, and hot passing electrons (Fig.\,\ref{fig:domain-schematic}).

%The timescales and frequencies of note are: i) the bounce frequency of a trapped electron $\nu_B$, ii) the inverse transit time of a passing ambient electron
%\begin{equation}
%    \nu_T = \frac{v_{T_a}}{L_p},
%\end{equation}
%where $v_{T_a} = \sqrt{2T_a/m_e}$ is the ambient electron thermal velocity, iii) the collision frequency of an electron with the plasmoid electrons
%\begin{equation}
%    \nu_p(v) = \frac{n_p e^4\ln\Lambda}{4\pi\varepsilon_0^2 m_e^2 v^3}
%\end{equation}
%for $n_p$ the plasmoid density and $v$ the velocity of the electron, iv) the heating time 
%\begin{equation}
   % \nu_h = \frac{n_a %e^4\ln\Lambda}{6\sqrt{2}\pi^\frac{3}{2}\varepsilon_0^2 %m_e^\frac{1}{2}T_a^\frac{3}{2}}, \label{eq:nu0}
%\end{equation}
%and v) the 
%. $\nu_h$ may be considered the frequency with which a trapped electron collides with the hot ambient electrons. For electrons near the trapped-passing separatrix (regions II and III) we find that $\nu_B \sim \nu_T \sim 10^7\,\mathrm{s}^{-1}$, $\nu_h \sim 10^{4}\,\mathrm{s}^{-1}$, and $\nu_p(v_{T_a}) \sim 10^{6}\,\mathrm{s}^{-1}$ with $n_a = 5\times10^{19}\,\mathrm{m}^{-3}$, $T_a = 5\,\mathrm{keV}$, $L_p = 1\,\mathrm{m}$, $N_p = 10^{22}\,\mathrm{m}^{-2}$ (implying $n_p = 10^{22}\,\mathrm{m}^{-3}$ and $\phi_m = 5\,\mathrm{kV}$).

In each region we employ the separation of timescales appropriate for the plasmoid and plasma parameters mentioned earlier in this section in order to obtain a simplified kinetic equations for the electrons. We find that trapped electrons collide with the cold, dense plasmoid electrons (and the plasmoid ions) much more quickly than with the passing electrons. At the same time, owing to the high temperature of the ambient plasma, the mean free path of passing and hot trapped electrons exceeds the length of the plasmoid; the plasmoid appears essentially transparent to the ambient electrons, and hot trapped electrons bounce inside the well many times before colliding. The latter effect means that trapped electrons behave `adiabatically' as the potential well expands.

%Although the values of $\nu_T$ and $\nu_p$ will change as the plasmoid expands, their relative values will remain approximately constant since both are inversely proportional to the plasmoid size $L_p$. $\nu_h$ as defined in Eq.\,\eqref{eq:nu0} is constant, but we note that the effective heating timescale will increase as the plasmoid heats up; one may multiply $\nu_h$ in Eq.\,\eqref{eq:nu0} by $(T-T_a)/T_a$ to obtain the inverse effective heating timescale. Hence it is a reasonable assumption that the relative values of $\nu_h$, $\nu_T$, and $\nu_p$ remain approximately constant throughout much of the expansion. The fact that $\nu_h$ and $\nu_p(v_{T_a})$ are separated by several orders of magnitude while $\nu_B$ and $\nu_p(v_{T_a})$ are only separated by one will be exploited to yield a greatly simplify the electron kinetic problem in region II.

We will show that, except at the very earliest stage of expansion, the ordering of timescales leads to the electrons reaching a `quasi-equilibrium' (QE) state which is characterised by rapid electron collisions within the plasmoid causing the electron distribution to exhibit a steady-state on the timescale on which trapped electrons collide with the plasmoid. The steady-state is established with no net flux of electrons into the trapped region of phase-space to prevent the `charging up' of the plasmoid (and hence the violation of quasineutrality) on this timescale. The QE electron distribution function is analogous to an equilibrium distribution function, which is indeed attained if electron-electron collisions are \textit{the} fastest effect. In our case, however, the bounce period of hot trapped electrons is considerably shorter than the collision timescale. We note that the QE state and the equilibrium state (which has a Maxwellian energy distribution) differ conceptually; a Maxwellian distribution exhibits no collisional flux, but the QE state is characterised by a vanishing \textit{divergence} of collisional flux; in this sense QE is a `dynamical' steady-state.

It will be shown that the QE distribution is specified in terms of the `deeply trapped' distribution function occupying region I, a Maxwellian with homogeneous temperature $T$, which is uniquely defined by two parameters. These two parameters must be such that there is no net flux of electrons into the trapped region on the timescales on which QE is established. This allows us to express one parameter of the lowest-order distribution in terms of the other. Once the electron distribution is known in terms of this remaining parameter, which we choose to be the temperature, its zeroth moment may be taken to obtain an expression for electron density, which, combined with the quasineutrality condition provides an implicit expression for the self-consistent electric potential $\phi$ in terms of the temperature. The velocity moment corresponding to line-integrated energy density is then taken over the electron kinetic equation to obtain an energy conservation law, which is practically used as the evolution equation for $T$.

A description of the expansion requires a model for ion motion. Two models were considered: a cold-fluid model with a single flow velocity and a collisionless kinetic (Vlasov) model. The first model is pragmatically justified by a possible application of this work being to provide a simplified model for pellet plasmoid expansion in an established fluid code. The second is justified by the long mean-free-path of hot ambient ions. These models represent opposite collisionality regimes for ions, and we therefore expect the shared qualitative properties to remain in a more sophisticated and accurate model for the ions. The qualitative property of greatest concern is the electron to ion energy transfer during the expansion.

With each ion model the system was evolved until the plasmoid and ambient densities were similar and the plasmoid electron temperature $T$ had reached $T_a$; the plasmoid assimilated with the ambient plasma. After this point the electric field does not provide much energy to the ions, so the energy transfer from electrons to ions may be considered complete. With the given choice of plasma parameters the densities and temperatures equilibrate at approximately the same time. With a larger line-integrated density $N_p$ the temperature equilibration would occur well before the densities are comparable. With a much smaller line-integrated density the densities become similar well before the temperatures have equilibrated. More discussion of how the QE formalism fits into the larger topic of plasmoid expansion is given in a later section.

Late in the plasmoid expansion with the cold-fluid model for ions, a steepening of the density profile results from the fast plasmoid moving into the ambient plasma causing a shock near the extremities of the plasmoid. This shock which will cause the generation of sound waves and solitons that will propagate into the ambient plasma. The wave propagation will sap the kinetic energy of the plasmoid and possibly the electron-ion energy balance. However, the shock and wave dynamics may only be properly accounted for with Poisson's equation for the electric potential since a deviation from quasineutrality will occur near the shock. As we neglect any deviation from quasineutrality, sound wave and soliton generation is suppressed.

In the Vlasov ion model, which accounts for the ambient ion temperature, no shock is observed; the density profile smoothly decreases to the ambient density. This is because many ambient ions can now traverse the entire plasmoid or are rapidly reflected from the potential, hence do not pile up at the expanding edge of the plasmoid. It is expected that as the ion temperature decreases the system is more prone to forming a shock.

\subsection{Self-similar solution to plasmoid expansion}
\cite{Aleynikov2019} provided the self-similar solution to plasmoid expansion given that the plasmoid is transparent to the ambient plasma. Although we seek to rectify the issues in the model therein, the density and temperature profiles obtained with the plasma parameters given earlier will be used to justify the ordering which will be used to simplify the electron kinetic problem. We require only order-of-magnitude estimates to find this ordering, so we deem the self-similar solution good enough for this purpose. However, as we wish to model the long-term expansion of the plasmoid, well past the applicability of the self-similar model, we must modify the profiles in \cite{Aleynikov2019} so that they are valid (i.e. giving a correct order of magnitude estimate) for the long-term expansion.

Firstly, we note that in the self-similar expansion the plasmoid electron temperature is given by $T = \nu_h T_a t$ for $t \ll \nu_h^{-1}$, where
\begin{equation}
    \nu_h = \frac{n_a e^4 \ln\Lambda}{6\sqrt{2}\pi^\frac{3}{2}\varepsilon_0^2 m_e^\frac{1}{2}T_a^\frac{3}{2}} \label{eq:nu_h}
\end{equation}
is the inverse heating time of the cold plasmoid electrons by a hot population of density $n_a$ and temperature $T_a$. The expression
\begin{equation}
    T \sim T_a\left(1 - \mathrm{e}^{-\nu_h t}\right) \label{eq:T_Aleynikov_modified}
\end{equation}
agrees with this linearly increasing temperature at early times, but exponentially approaches $T_a$ as time advances, which is the characteristic behaviour of a cold Maxwellian being heated by a hotter one. Therefore we expect the above expression to be adequate in describing the plasmoid electron temperature in both the short and long-term evolution of the plasmoid.

In the self-similar solution, the density becomes infinite as $t\rightarrow0$ and vanishes for $t\rightarrow\infty$, neglecting the fact that the electron density approaches $n_a$ as time advances. Therefore the expression
\begin{equation}
    n_m \sim N_p \nu_h\sqrt{\frac{3 m_i}{8\pi (\nu_h t)^3 T_a}} + n_a, \label{eq:n_m_Aleynikov_modified}
\end{equation}
for the peak plasma density, which simply adds the ambient electron density $n_a$ to the self-similar solution, is a plausible expression for both long and short term evolution.

Although the electric potential in \cite{Aleynikov2019} diverges as $|z| \rightarrow \infty$, the Boltzmann relation
\begin{equation}
    e\phi_m \sim T \ln\left(\frac{n_m}{n_a}\right) \label{eq:Boltzmann}
\end{equation}
provides a good estimate for the height of the potential. This estimate is also supported by solution to the self-consistent electron kinetic problem in \cite{Arnold2023}, which showed that the potential height is the same order of magnitude as that suggested by the Boltzmann relation, its exact value being somewhat larger when $T \ll T_a$.

Equation \eqref{eq:n_m_Aleynikov_modified} expresses is the \textit{peak} plasma density, which we will subsequently use in the ordering. Of course, electrons move throughout the plasmoid, more for passing electrons and less for trapped electrons, so the most rigorous approach would be to consider `average' quantities throughout the orbit. This would be very complicated, and relies on a detailed knowledge of the shape of the potential which we have not yet obtained. Therefore, we apply the ordering using the quantities at the peak of the plasmoid with the reasoning that the plasmoid density is very large at its peak and decreases rapidly as one moves away from it. Hence, when considering trapped electrons the orbit-average of any quantity will be heavily weighted by the value at the peak. When consider passing electrons, we will actually obtain the same expressions as those obtained by rigorous consideration.

%\begin{figure}
%    \centering
%    \includegraphics[width=0.8\textwidth]{f-schematic.pdf}
%    \caption{Schematic of the electron distribution $f$ inside a plasmoid as a function of parallel energy $\mathcal{E}_\parallel = m_e v_\parallel^2/2 -e \phi$ (cf. \cite{Arnold2023}, Figs.\,2,3,4,5 with $\mathcal{E} = \mathcal{E}_\parallel$).}
%    \label{fig:f-schematic}
%\end{figure}

\FloatBarrier

\section{Electron kinetics}
The kinetic equation for the electron distribution function $f$ is given by
\begin{equation}
        \frac{\partial f}{\partial t} + v_\parallel \frac{\partial f}{\partial
        z} + \frac{e}{m_e}\frac{\partial \phi}{\partial z}\frac{\partial f}{\partial
        v_\parallel} = C(f,f) + \sum_k C_{e,ik}(f) \label{eq:kinetic_1}
\end{equation}
in the variables $(v_\parallel, v_\perp, z, t)$, where $v_\parallel$ is the velocity parallel to the field line, $v_\perp$ is the speed perpendicular to the field line, $z$ is the coordinate along the field line, $t$ is time, $C(f,f)$ is the electron self-collision operator and $C_{e,ik}(f)$ is the collision operator for electrons colliding against ion species $k$.

We change to the independent variables $(\mathcal{E}_\parallel =
m_ev_\parallel^2/2 - e\phi,\mathcal{E}_\perp = m_e v_\perp^2/2,z,t)$:
\begin{equation}
        \frac{\partial f}{\partial t} + v_\parallel \frac{\partial f}{\partial
        z} - e \frac{\partial
        \phi}{\partial t}\frac{\partial f}{\partial \mathcal{E}_\parallel} = C(f,f) + \sum_k C_{e,ik}(f), \label{eq:kinetic_2}
\end{equation}
showing that collisionless change in electron energy is associated with
time-variation of the electric potential. We note that with a stationary potential both $\mathcal{E}_\parallel$ and $\mathcal{E}_\perp$
are constants of motion in the absence of collisions. Passing electrons have $\mathcal{E}_\parallel \ge 0$ and trapped $\mathcal{E}_\parallel < 0$. The minimum parallel energy of an electron is $\mathcal{E}_\parallel = -e\phi_m$, which corresponds to an electron with $v_\parallel = 0$ which remains at $z = 0$.

We now solve the kinetic equations in regions I, II, and III. We split up the distribution function into its representation in each region, $f_\mathrm{I}$, $f_\mathrm{II}$, and $f_\mathrm{III}$. That is, when we are in region I, for example, we solve the kinetic equation for $f_\mathrm{I}$. No matter where we are in phase-space, collisions are experienced with every other region of phase-space via $C(\cdot, f_\mathrm{I} + f_\mathrm{II} + f_\mathrm{III})$. In this sense it is understood that $f_\mathrm{I}$ (or $f_\mathrm{II}$ or $f_\mathrm{III}$) is the entire distribution function in region I (or II or III), but is zero outside of its own region.

\subsection{Solving the kinetic equation for passing electrons (region III)}
The kinetic equation for electrons in region III is given by
\begin{equation}
    \frac{\partial f_\mathrm{III}}{\partial t} + v_\parallel \frac{\partial f_\mathrm{III}}{\partial
        z} - e \frac{\partial
        \phi}{\partial t}\frac{\partial f_\mathrm{III}}{\partial \mathcal{E}_\parallel} = C(f_\mathrm{III},f) + \sum_k C_{e,ik}(f_\mathrm{III}). \label{eq:kinetic_region_III}
\end{equation}
The collision frequency of an electron inside the plasmoid with the plasmoid  electrons and ions is approximately given by
\begin{equation}
    \nu_p(v) = \frac{n_p(1 + Z_\mathrm{eff})e^4 \ln\Lambda}{8\pi\varepsilon_0^2 m_e^2 v^3}, \label{eq:nu_p}
\end{equation}
where $v$ is the electron speed, $n_p$ is the plasmoid electron density, and
\begin{equation}
    Z_\mathrm{eff} = \frac{\sum_k n_{ik}Z_k^2}{\sum_k n_{ik}Z_k}
\end{equation}
is the effective charge of the ions. Equation \eqref{eq:nu_p} is derived from the frequency with which an electron experiences pitch-angle scattering, assuming that quasineutrality
\begin{equation}
    n_e \approx n_p = \sum_k n_{ik} Z_k,
\end{equation}
where $n_e$ is the electron density, holds  inside the plasmoid. The typical velocity of a passing electron is given by the ambient electron thermal velocity $v_{T_a} = \sqrt{2T_a/m_e}$. The inverse of the time taken for a passing electron to transit the plasmoid is $\nu_T = v_{T_a}/L_p$ for plasmoid length $L_p = N_p/n_p(z = 0)$. Hence, for a hydrogenic plasma, in region III the ratio of the collision frequency with the plasmoid and the inverse transit time is given by
\begin{equation}
    \frac{\nu_p(v_{T_a})}{\nu_T} := \mu = \frac{N_p e^4 \ln\Lambda}{16\pi\varepsilon_0^2 T_a^2},.\label{eq:opacity}
\end{equation}
With the plasma and plasmoid parameters given in the Introduction this ratio is much less than unity:
\begin{equation}
    \mu \ll 1. \label{eq:nu_p_over_nu_T_small}
\end{equation} 
where for the purpose of calculation the Coulomb logarithm $\ln\Lambda$ is assumed to be equal to $15$. This implies that the mean free path of passing electrons is much longer than the plasmoid, so the plasmoid appears essentially transparent to ambient electrons. We therefore refer to $\mu$ as the opacity of the plasmoid. $\mu$ is independent of any parameters that change during the expansion, hence it is small throughout the \textit{entire} expansion.

The terms in the kinetic equation \eqref{eq:kinetic_region_III} containing time derivatives correspond to collisionless changes in the energy of passing electrons, which certainly occur on a longer timescale than the transit time. Therefore the shortest timescale in the region III is the transit time, which is associated with the convective term in the kinetic equation; the lowest-order kinetic equation for passing electrons is
\begin{equation}
    v_\parallel \frac{\partial f_\mathrm{III}}{\partial z} = 0,
\end{equation}
which implies that $f_\mathrm{III}$ is independent of $z$.
The kinetic equation for passing electrons \eqref{eq:kinetic_region_III} then reduces to
\begin{equation}
     \frac{\partial f_\mathrm{III}}{\partial t} - e \frac{\partial
        \phi}{\partial t}\frac{\partial f_\mathrm{III}}{\partial \mathcal{E}_\parallel} = C(f_\mathrm{III},f) + \sum_k C_{e,ik}(f_\mathrm{III}), \label{eq:kinetic_passing_unaveraged}
\end{equation}
the solution to which can be obtained immediately by bounce-averaging. We define the bounce integral of a function $g(\mathcal{E}_\parallel,\mathcal{E}_\perp,z,t)$ to be
\begin{equation}
    \oint g\,\mathrm{d}z := 2 \int^{z_c}_{-z_c}g\,\mathrm{d}z,
\end{equation}
where $z_c(\mathcal{E}_\parallel,t) > 0$ is the turning point such that
\begin{equation}
    \mathcal{E}_\parallel + e\phi(z_c,t) = 0
\end{equation}
(cf. Fig.\,\ref{fig:well_schematic}). The bounce average of $g$ is given by
\begin{equation}
    \left<g\right> = \frac{1}{\tau}\oint g\,\mathrm{d}z 
\end{equation}
for bounce period
\begin{equation}
    \tau = \oint \frac{\mathrm{d}z}{v_\parallel}.
\end{equation}
We note that on an infinitely long magnetic field line the bounce-average of any function $g$ for $\mathcal{E}_\parallel > 0$ is given by
\begin{equation}
    \left<g\right>\big|_{\mathcal{E}_\parallel > 0} = \lim_{|z| \rightarrow \infty} g.
\end{equation}
Since the potential vanishes at infinity and we expect the distribution function to be constant at infinity, the bounce average of Eq\,\eqref{eq:kinetic_passing_unaveraged} is solved by the Maxwellian defining the ambient plasma:
\begin{equation}
        f_\mathrm{III} = f_a = n_a \left(\frac{m_e}{2\pi
        T_a}\right)^\frac{3}{2}\mathrm{e}^{-\frac{\mathcal{E}}{T_a}}. \label{eq:f_III}
\end{equation}

We note that, as mentioned earlier, we have used the expression for peak plasmoid density to deduce the ordering \eqref{eq:nu_p_over_nu_T_small} and obtain Eq.\,\eqref{eq:f_III}, despite the fact that plasmoid electrons spend little time inside the plasmoid. The same result can be obtained by a slightly different approach: Eq.\,\eqref{eq:opacity} is identical to the expression for the number of passing electrons slowed down by collisions with the plasmoid as they traverse it \citep{Arnold2021}. If the number of passing electrons slowed by the plasmoid is small, then the passing electron distribution is modified little by collisions and is simply replenished from flux emerging from $|z| \rightarrow \infty$, resulting in the distribution identical to that at infnity; hence Eq.\,\eqref{eq:f_III}.

\subsection{Solving the kinetic equation for deeply trapped electrons (region I)}
The kinetic equation for electrons in region I is given by
\begin{equation}
    \frac{\partial f_\mathrm{I}}{\partial t} + v_\parallel \frac{\partial f_\mathrm{I}}{\partial
        z} - e \frac{\partial
        \phi}{\partial t}\frac{\partial f_\mathrm{I}}{\partial \mathcal{E}_\parallel} = C(f_\mathrm{I},f_\mathrm{I}) + C(f_\mathrm{I},f_\mathrm{II} + f_\mathrm{III}) + \sum_k C_{e,ik}(f_\mathrm{I}). \label{eq:kinetic_region_I}
\end{equation}
We expect the potential well to be parabolic at its peak; deeply trapped electrons bounce inside this parabola and collide with the plasmoid when its density is near its peak. For a parabolic potential well of height $\phi_m$ and width $L_p$, the bounce frequency is given by
\begin{equation}
    \nu_B \sim \frac{L_p}{\sqrt{2e\phi_m/m_e}},\label{eq:nu_B}
\end{equation}
and is associated with the convective term, so we write
\begin{equation}
    v_\parallel \frac{\partial f_\mathrm{I}}{\partial z} \sim \nu_B f_\mathrm{I}.
\end{equation}

Substituting temperature \eqref{eq:T_Aleynikov_modified} and density \eqref{eq:n_m_Aleynikov_modified} from the modified self-similar solution into the Boltzmann relation \eqref{eq:Boltzmann} yields
\begin{equation}
    \frac{e\phi_m}{T_a} \sim (1-\mathrm{e}^{-\nu_h t})\ln\left(1 + \frac{2}{\pi\sqrt{3}} \sqrt{\frac{m_i}{m_e}} \mu  (\nu_h t)^{-\frac{3}{2}}\right).
\end{equation}
Given that $(2/(\pi\sqrt{3}))\sqrt{m_i/m_e}\mu$ is at least order unity, the above is order unity for several heating times $\nu_h^{-1}$, with the exception of the very early stage of the expansion. This is the case for the plasma parameters given in the Introduction. We also note that while $T < T_a$ the Boltzmann relation provides something of an underestimate, strengthening the argument that $e\phi_m \sim T_a$ for smaller values of $\mu$.

Owing to the height of the potential, the bounce frequency in region I is of the same order as the transit frequency of region III:
\begin{equation}
    \nu_B \sim \frac{L_p}{\sqrt{2T_a/m_e}}.\label{eq:nu_B_2}
\end{equation}

Since we expect $f_\mathrm{I}$ to correspond to a dense population of cold electrons, we associate the collision terms against $f_\mathrm{I}$ with the frequency of collisions with the plasmoid; we write
\begin{equation}
    C(f_\mathrm{I},f_\mathrm{I}) \sim \sum_k C_{e,ik}(f_\mathrm{I}) \sim \nu_p\left(v_T\right) f_\mathrm{I},
\end{equation}
where $v_T = \sqrt{2T/m_e}$ the typical electron speed within region I. 

Since $f_\mathrm{II}$ and $f_\mathrm{III}$ represent the hot tail, we associate the collision terms against these with the heating rate. Since $T$ approaches exponentially $T_a$ as time advances, this heating rate decreases exponentially as time advances: we define the heating rate to be
\begin{equation}
    \frac{1}{T_a}\frac{\mathrm{d} T}{\mathrm{d} t} = \nu_h \mathrm{e}^{-\nu_h t}
\end{equation}
and write
\begin{equation}
    C(f_\mathrm{I},f_\mathrm{II} + f_\mathrm{III}) \sim \nu_h \mathrm{e}^{-\nu_h t} f_\mathrm{I}.
\end{equation}

Comparing the frequency of collisions with the plasmoid with the heating using the modified self-similar temperature \eqref{eq:T_Aleynikov_modified} and density \eqref{eq:n_m_Aleynikov_modified} gives
\begin{equation}
    \frac{\nu_h \mathrm{e}^{-\nu_h t}} {\nu_p(v_T)} \sim \frac{4}{3\sqrt{\pi}}\mathrm{e}^{-\nu_h t}\left(1 - \mathrm{e}^{-\nu_h t}\right)^\frac{3}{2}\left(1 + \frac{2}{\pi\sqrt{3}}\sqrt{\frac{m_i}{m_e}}\mu (\nu_h t)^{-\frac{3}{2}}\right)^{-1}.
\end{equation}
With the plasma parameters used in the Introduction the above is much smaller than unity for all times: collisions with the plasmoid occur much more frequently than collisions that cause heating. This effect is also enhanced by the fact that in a parabolic potential well we expect the heating rate to be slightly reduced, due to the reduction in the density of passing electrons and their reduction in collisionality \citep{Arnold2023}. Hence we can write
\begin{equation}
    \frac{\nu_h \mathrm{e}^{-\nu_h t}} {\nu_p(v_T)} \ll 1
\end{equation}

Since region I represents the cold electrons,
we expect the time-dependent terms in Eq.\,\eqref{eq:kinetic_region_I} to act on a much longer timescale than the collision time. In region III we noted that the transit frequency greatly exceeds the collision frequency with the plasmoid. However, as we move into region I, which contains lower-velocity electrons, the bounce frequency (which is comparable to the transit frequency, cf. Eq.\,\eqref{eq:nu_B_2}), remains the same as the collision frequency increases. Therefore, collisions with the plasmoid and bounce motion are associated with the two shortest timescales in region I.
Accordingly, the lowest-order kinetic equation in region I is
\begin{equation}
    v_\parallel \frac{\partial f_\mathrm{I}}{\partial z} = C(f_\mathrm{I},f_\mathrm{I}) + \sum_k C_{e,ik}(f_\mathrm{I}). \label{eq:kinetic_region_I_Maxwellian}
\end{equation}
When $T \ll T_a$, $T \ll e\phi_m$, which allows us to define $\mathcal{E}_\mathrm{I/II}$  such that region I extends for several times $T$ in both the parallel and perpendicular direction. Therefore, the solution to the above, assuming that collisions with ions are well-approximated by pitch-angle scattering, is a Maxwellian in energy
\begin{equation}
    f_\mathrm{I} = f_0 = \eta \left(\frac{m_e}{2\pi T}\right)\mathrm{e}^{-\frac{\mathcal{E}}{T}} \label{eq:f_I}
\end{equation}
for parameters $\eta(t)$, $T(t)$ \citep{Aleynikov2019}. We see now that in the earlier work \cite{Aleynikov2019,Runov2021,Arnold2021} only electrons in region I, where a purely Maxwellian electron distribution function is exhibited, are treated, whereas in this investigation we continue the analysis in regions II and III.

The remaining kinetic equation in region I, corresponding to the heating timescale, is given by
\begin{equation}
    \frac{\partial f_\mathrm{I}}{\partial t} - e \frac{\partial
        \phi}{\partial t}\frac{\partial f_\mathrm{I}}{\partial \mathcal{E}_\parallel} = C(f_\mathrm{I},f_\mathrm{II} + f_\mathrm{III}), \label{eq:kinetic_region_I_higher}
\end{equation}
which captures the collisionless change in electron energy due to the expanding well and the heating of the cold Maxwellian by the hot electrons.

\subsection{Choosing $\mathcal{E}_\mathrm{I/II}$}
Owing to the ordering, when $T \ll T_a$ we understand that the potential well is deep enough for a Maxwellian to reside in region I, provided $\mathcal{E}_\mathrm{I/II}$ is chosen close enough to zero. We now decide upon an explicit definition for $\mathcal{E}_\mathrm{I/II}$ which is consistent with the distribution in region I: we require that collisions with the cut-off Maxwellian in region I are well-approximated by collisions with the full Maxwellian that extends to arbitrarily large energies. This will allow the linearisation of the kinetic problem in region II in terms of collisions with the Maxwellian. However, we must not artificially extend region I past the point where the distribution function would be Maxwellian; this would result in an incorrect distribution function.

From the Boltzmann relation \eqref{eq:Boltzmann} and the density of the full Maxwellian distribution \eqref{eq:f_I} being $n_p = \eta \exp(e\phi/T)$ we find that
\begin{equation}
    n_p \sim n_a \mathrm{e}^\frac{e\phi}{T}.
\end{equation}
Considering an electron with with parallel energy $\mathcal{E}_\mathrm{I/II}$, it has a turning point $z_c$ such that $e\phi(z_c) + \mathcal{E}_\mathrm{I/II} = 0$. Therefore, at this turning point,
\begin{equation}
    n_p(z_c) \sim n_a \mathrm{e}^{-\frac{\mathcal{E}_\mathrm{I/II}}{T}}.
\end{equation}
During its orbit, this electron collides with a plasmoid density that strictly larger than the above. Therefore, if we choose $\mathcal{E}_\mathrm{I/II}$ such that $n_p(z_c) > a n_a$ for $a \gg 1$, then the collisions the electron experiences are completely dominated by collisions with the cold Maxwellian \textit{throughout its entire orbit}. Above this energy, the electron collides considerably with the ambient electrons in the extremities of the plasmoid \textit{as well as with the plasmoid electrons in the core}, so the distribution function at these higher energies is not necessarily Maxwellian. Hence, the upper bound for $\mathcal{E}_\mathrm{I/II}$ is expressed as
\begin{equation}
    \mathcal{E}_\mathrm{I/II} < -T\ln a.
\end{equation}

The lower bound is fixed by collisions with the cut-off Maxwellian in region I being well-approximated by collisions with the full Maxwellian. The simplest way to guarantee this is to have
\begin{equation}
    \frac{f_0(\mathcal{E}_\mathrm{I/II})}{f_0(-e\phi_m)} < \frac{1}{a}
\end{equation}
for $a \gg 1$. Then, the lower bound for $\mathcal{E}_\mathrm{I/II}$ is given by
\begin{equation}
    \mathcal{E}_\mathrm{I/II} > -e\phi_m + T\ln a.
\end{equation}

\subsection{Deriving the kinetic equation for hot trapped electrons (region II)}
The kinetic equation in region II is given by
\begin{equation}
    \frac{\partial f_\mathrm{II}}{\partial t} + v_\parallel \frac{\partial f_\mathrm{II}}{\partial
        z} - e \frac{\partial
        \phi}{\partial t}\frac{\partial f_\mathrm{II}}{\partial \mathcal{E}_\parallel} = C(f_\mathrm{II},f_\mathrm{I}) + C(f_\mathrm{II},f_\mathrm{II} + f_\mathrm{III}) + \sum_k C_{e,ik}(f_\mathrm{II}). \label{eq:kinetic_region_II}
\end{equation}

As the intermediate region, the ordering is most complex for the hot trapped electrons. More care must also be taken when considering terms with time derivatives as the collision frequency is lower in region II than region I. The typical velocity in region I is of order $\sqrt{2e\phi_m/m_e} \sim v_{T_a}$, so the collision frequency with the plasmoid in region II is of the same order as in region III. In region II the bounce frequency is of the same order as in region I (Eq.\,\eqref{eq:nu_B_2}). Hence we write
\begin{equation}
    C(f_\mathrm{II},f_\mathrm{I}) \sim \sum_k C_{e,ik}(f_\mathrm{I}) \sim \nu_p(v_{T_a})f_\mathrm{II},
\end{equation}
\begin{equation}
    C(f_\mathrm{II},f_\mathrm{II} + f_\mathrm{III}) \sim \nu_h \mathrm{e}^{-\nu_h t}f_\mathrm{II},
\end{equation}
\begin{equation}
    v_\parallel \frac{\partial f_\mathrm{II}}{\partial z} \sim \nu_B f_\mathrm{II},
\end{equation}
noting that both the collision frequency with the plasmoid and the heating rate are much smaller than the bounce frequency:
\begin{equation}
    \frac{\nu_p(v_{T_a})}{\nu_B} = \mu \ll 1,
\end{equation}
\begin{equation}
    \frac{\nu_h \mathrm{e}^{-\nu_h t}}{\nu_B} = \frac{4}{3\sqrt{\pi}}\frac{n_a L_p \mathrm{e}^{-\nu_h t}}{N_p} \mu \ll 1.
\end{equation}
As in region I, we assume that the terms containing time derivatives correspond to a timescale much longer than the bounce period.

Then, the shortest timescale in the system is the bounce period, which leads to the lowest-order equation
\begin{equation}
    v_\parallel \frac{\partial f_\mathrm{II}}{\partial z} = 0,
\end{equation}
meaning that $f_\mathrm{II}$ (and hence $f$ as a whole) is independent of $z$. The higher-order kinetic equation can then be bounce-averaged, yielding
\begin{equation}
    \frac{\partial f_\mathrm{II}}{\partial t} - \frac{1}{\tau}\frac{\partial J}{\partial t}\frac{\partial f_\mathrm{II}}{\partial \mathcal{E}_\parallel} = \left<C(f_\mathrm{II},f_\mathrm{I}) + \sum_k C_{e,ik}(f_\mathrm{II})\right> + \left<C(f_\mathrm{II},f_\mathrm{II} + f_\mathrm{III})\right> \label{eq:kinetic_f_region_II_ba}
\end{equation}
where 
\begin{equation}
    J(\mathcal{E}_\parallel,t) = \oint m_e v_\parallel\,\mathrm{d}z =
        \sqrt{2m_e}\oint \sqrt{\mathcal{E}_\parallel + e\phi}\,\mathrm{d}z
\end{equation}
is the second adiabatic invariant for an electron bouncing in the well.

Now we analyse the timescales on which the time-dependent terms act and compare them to other timescales. Since $f_\mathrm{II}$ is equal to $f_0$ at $\mathcal{E} = \mathcal{E}_\mathrm{I/II}$ and equal to $f_a$ at $\mathcal{E}_\parallel = 0$ it serves to perform this analysis at each boundary. At the trapped-passing separatrix, $\mathcal{E}_\parallel = 0$, $f_\mathrm{II}$ must be equal to $f_a$, which is constant in time; the first term on the left hand side of Eq.\,\eqref{eq:kinetic_f_region_II_ba} vanishes. The second term represents the adiabatic change in electron energy as the well expansion, which in \cite{Aleynikov2019} was shown to occur on the heating timescale:
\begin{equation}
    - \frac{1}{\tau}\frac{\partial J}{\partial t}\frac{\partial f_\mathrm{II}}{\partial \mathcal{E}_\parallel} \sim \nu_h \mathrm{e}^{-\nu_h t} f_\mathrm{II}.\label{eq:adiabatic_ordering}
\end{equation}
At $\mathcal{E} = \mathcal{E}_\mathrm{I/II}$ Eq.\,\eqref{eq:adiabatic_ordering} also holds. However, since the Maxwellian in region I has a temperature that changes in time, the time derivative of $f_\mathrm{II}$ is not zero here:
\begin{equation}
    \frac{\partial f_\mathrm{II}}{\partial t}\bigg|_{\mathcal{E} = \mathcal{E}_\mathrm{I/II}} = \left[\frac{1}{\eta}\frac{\mathrm{d} \eta}{\mathrm{d} t} + \left(\frac{\mathcal{E}_\mathrm{I/II}}{T} - \frac{3}{2}\right)\frac{1}{T}\frac{\mathrm{d} T}{\mathrm{d} t}\right] f_\mathrm{II}\bigg|_{\mathcal{E} = \mathcal{E}_\mathrm{I/II}}. \label{eq:time_dependent_region_I/II}
\end{equation}

We can choose $|\mathcal{E}_\mathrm{I/II}| \sim T$, and we see that $(\partial \eta/\partial t)/\eta \sim (\partial T/\partial t)/T$, so Eq.\,\eqref{eq:time_dependent_region_I/II} can be approximated by
\begin{equation}
    \frac{\partial f_\mathrm{II}}{\partial t}\bigg|_{\mathcal{E} = \mathcal{E}_\mathrm{I/II}} \sim \frac{1}{T}\frac{\mathrm{d} T}{\mathrm{d} t}f_\mathrm{II}\bigg|_{\mathcal{E} = \mathcal{E}_\mathrm{I/II}}.
\end{equation}
That is, the timescale on which the term acts, which we define via the frequency
\begin{equation}
    \nu_t = \frac{1}{T}\frac{\mathrm{d} T}{\mathrm{d} t} = \frac{T_a}{T} \nu_h \mathrm{e}^{-\nu_h t} ,
\end{equation}
is the time taken for $T$ to increase by a factor of $\mathrm{e}$. When $T$ is small this can be a very short time, so $\nu_t$ cannot simply be assumed to be small compared to the collision time.

The profiles from the modified self-similar solution give
\begin{equation}
    \frac{\nu_t} {\nu_p(v_{T_a})} \sim \frac{4}{3\sqrt{\pi}}\mathrm{e}^{-\nu_h t}(1 - \mathrm{e}^{-\nu_h t})^\frac{1}{2}\left(1 + \frac{2}{\pi\sqrt{3}}\sqrt{\frac{m_i}{m_e}}(\nu_h t)^{-\frac{3}{2}}\right)^{-1},
\end{equation}
which is always much less than unity for the plasma parameters given in the Introduction; we write
\begin{equation}
    \frac{\nu_t} {\nu_p(v_{T_a})} \ll 1.
\end{equation}

So, the collision term with the with the plasmoid corresponds to the shortest timescale in Eq.\,\eqref{eq:kinetic_f_region_II_ba}; the lowest-order kinetic equation is therefore
\begin{equation}
    \left<C(f_\mathrm{II},f_\mathrm{I}) + \sum_k C_{e,ik}(f_\mathrm{II})\right> = 0.\label{eq:kinetic_region_II_QE}
\end{equation}
The distribution function must continuous in collisional kinetic problems, hence Eq.\,\eqref{eq:kinetic_region_II_QE} must be solved with boundary conditions ensuring continuity:
\begin{equation}
    f_\mathrm{II}(\mathcal{E} = \mathcal{E}_\mathrm{I/II}) = f_0(\mathcal{E} = \mathcal{E}_\mathrm{I/II}), \label{eq:QE_BC_2}
\end{equation}
\begin{equation}
        f_\mathrm{II}(\mathcal{E}_\parallel = 0) = f_a(\mathcal{E}_\parallel = 0).
        \label{eq:QE_BC_1}
\end{equation}
The higher-order equation in region II is
\begin{equation}
    \frac{\partial f_\mathrm{II}}{\partial t} - \frac{1}{\tau}\frac{\partial J}{\partial t}\frac{\partial f_\mathrm{II}}{\partial \mathcal{E}_\parallel} = \left<C(f_\mathrm{II},f_\mathrm{II} + f_\mathrm{III})\right>, \label{eq:kinetic_region_II_ba_higher}
\end{equation}
which describes the heating and expansion timescales.

\subsection{The quasi-equilibrium problem}
The distribution function in region II is obtained by solving Eq.\,\eqref{eq:kinetic_region_II_QE}, which must be be supplemented by boundary conditions \eqref{eq:QE_BC_2},\eqref{eq:QE_BC_1} enforcing continuity of the distribution function into regions I and III. Conceptually, the kinetic problem in region II describes a \textit{quasi-equilibrium} (QE); hot trapped electrons experience rapid collisions against a Maxwellian (and are isotropised by collisions with ions), but the tail of the distribution is forced to meet a Maxwellian of a different temperature at the trapped-passing separatrix.

When $T \ll T_a$ collisions with the distribution in region I are well-approximated by collisions with the full Maxwellian: $C(\cdot,f_\mathrm{I}) \approx C(\cdot, f_0)$. Since it only remains to solve the kinetic problem in region II, it is unnecessary to have a subscript, so we write $f = f_\mathrm{II}$ in region II. Hence the QE kinetic equation can be written as
\begin{equation}
    \left<C_\mathrm{QE}(f)\right> = 0 \label{eq:QE}
\end{equation}
for
\begin{equation}
    C_\mathrm{QE}(f) = C(f,f_0) + \sum_k C_{e,ik}(f).
\end{equation}

Further, owing to the fact that collisions are linearised in terms of collisions against a full Maxwellian, the lower boundary condition Eq.\,\eqref{eq:QE_BC_2} can actually be applied at $\mathcal{E} = -e\phi_m$ rather than $\mathcal{E}_\mathrm{I/II}$.

\subsection{Range of validity of the ordering} \label{sec:validity}
The ordering developed in this section is based upon the self-similar expansion in \cite{Aleynikov2019}, modified to provide plausible profiles at the later stages of the expansion, given a line-integrated plasmoid density $N_p = 10^{22}\,\mathrm{m}^{-2}$ in an ambient plasma of density electron $n_a = 5\times10^{19}$ at a temperature of $5\,\mathrm{keV}$. The requirements of the ordering are that during most of the expansion the potential height is of order the ambient temperature:
\begin{equation}
    e\phi_m \sim T_a,
\end{equation}
that the plasmoid is transparent to passing and hot trapped electrons:
\begin{equation}
    \mu = \frac{\nu_p(v_{T_a})}{\nu_T} \sim \frac{\nu_p(v_{T_a})}{\nu_B} \ll 1,
\end{equation}
that the heating rate is much lower than the collision frequency with the plasmoid:
\begin{equation}
    \frac{\nu_h \mathrm{e}^{-\nu_h t}}{\nu_p(v_{T_a})} \ll 1,
\end{equation}
\begin{equation}
    \frac{\nu_h \mathrm{e}^{-\nu_h t}}{\nu_p(v_T)} \ll 1,
\end{equation}
and that the time taken for the plasmoid electron temperature to increase by a factor of $\mathrm{e}$ is much larger than the collision time:
\begin{equation}
    \frac{1}{T}\frac{\mathrm{d} T}{\mathrm{d} t} \ll \nu_p(v_{T_a}),
\end{equation}
\begin{equation}
    \frac{1}{T}\frac{\mathrm{d} T}{\mathrm{d} t} \ll \nu_p(v_T).
\end{equation}

The transparency of the plasmoid is dependent upon the line-integrated plasmoid density not being too large, and $e\phi_m$ being of order $T_a$ is dependent upon the line-integrated plasmoid density not being too small; satisfying both conditions does formally somewhat constrain the values of the line-integrated density.

However, the relative simplicity of the resulting kinetic problem provides strong motivation for using the formalism: we immediately obtain the distribution function in two out of three regions, and in the remaining region we must solve a steady-state kinetic equation. Then, the macroscopic expansion is described by a kinetic equation of which \textit{velocity moments} can be taken, in order to obtain much simpler time-dependent equations than one would by including the time-dependent terms directly in the kinetic problem. In this sense, the formalism is analogous to the Braginskii equations, but valid for systems with a long rather than short hot electron mean free path. 

We reiterate that the kinetic problem in region I and II was formally derived assuming $T \ll T_a$, which permitted a choice of $\mathcal{E}_\mathrm{I/II}$ that guaranteed a Maxwellian distribution in region I, collisions with which are well-approximated by collisions with the full Maxwellian. This ultimately allowed the linearisation of the kinetic problem in region II in terms of collisions with the full Mawwellian. However, we note that when $T = T_a$, the entire distribution function will be a single Maxwellian, \textit{yet} Eq.\,\eqref{eq:QE} \textit{would still be satisfied}. This is because electrons in region II would be colliding with Maxwellian electrons with the same temperature in regions I and III. We therefore conclude that the whole formalism is valid when $T \ll T_a$ \textit{and} when $T \rightarrow T_a$.

Therefore the formalism can actually accurately model the expansion \textit{outside} of its formal ordering (which has $T \ll T_a$), which is characteristic of robust simplifications of kinetic problems, such as the Braginskii equations, which often achieve a level of qualitative correctness even when the mean free path is long and the distribution function is not very close to a Maxwellian. By the same argument one could expect that the formalism here is qualitatively correct somewhat outside of the range of parameters that leads to the ordering.

As mentioned in the previous subsection, the boundary condition \eqref{eq:QE_BC_2} can be applied at $\mathcal{E} = -e\phi_m$ in the QE problem. This solves the problem of how to choose $\mathcal{E}_\mathrm{I/II}$ when $T$ approaches $T_a$ and the well becomes too shallow to contain several $T$: when solving the QE problem we can always choose $\mathcal{E}_\mathrm{I/II} = -e\phi_m$.

The formalism has been developed specifically with pellet plasmoids in mind, and essentially models plasmoid expansion with `intermediate' line-integrated densities. An alternate approach, which is more suited to the abstract study of plasmoid expansion, is to consider the limit as the line-integrated density goes to zero or infinity. Then, the ratio $e\phi_m/T_a$ is a large or small parameter on which an ordering may be based.

When the line-integrated density is very large, the plasmoid and ambient temperatures will equilibrate before the plasmoid density is comparable to the ambient density. Then, the expansion can be described simply with Maxwellian electrons from an early stage. If instead it is very small, the densities become comparable well before the temperatures have equilibrated. In our case, with an intermediate line-integrated density, the two occur at approximately the same time; certainly one cannot assume $T = T_a$ or $n_{pm} \approx n_a$ from an early stage.

When constructing the ordering, the opacity $\mu$ was given assuming that intra-species collisions dominate; ambient ions collide most quickly with plasmoid ions and ambient electrons collide most quickly with plasmoid electrons. This is the case when the thermal velocities of ions and electrons are disparate. However, if the thermal velocities of an ion population $k$ and an electron population are comparable, then the friction of the ions on the electrons is actually $m_{ik}/m_e$ times larger than the ion-ion collision frequency \citep{Helander2002}. The thermal velocities of the ambient ions and plasmoid electrons are actually comparable for a brief window of time where $T$ is extremely small. However, collisions of the ambient ions with plasmoid electrons in this regime cause rapid heating of the plasmoid electrons, therefore driving $T$ up and out of the regime where the plasmoid electrons and ambient ions have a comparable thermal velocity. Hence these collisions are negligible outside of the very early stages of the plasmoid expansion, where the ordering is, anyway, not satisfied; we restrict our attention to times later than this.

\subsection{Expressing the quasi-equilibrium equation in the variables $(\mathcal{E}_\parallel,\mathcal{E}_\perp)$}
The collision operator against $f_0$ in the variables $(v,\theta,z,t)$ for pitch-angle $\theta$ (assuming that $f$ is independent of the azimuthal angle of the velocity $\varphi$) is given by
\begin{equation}
\begin{split}
            C(f, f_0) = \frac{e^4 \ln\Lambda}{4\pi\varepsilon_0^2 m_e^2}\Bigg\{&\frac{1}{v^3}\frac{g^\prime(x)}{v_T}\frac{1}{2\sin\theta}\frac{\partial}{\partial \theta}\left(\sin\theta \frac{\partial f}{\partial \theta}\right) +\\ &\frac{1}{v^2}\frac{\partial}{\partial v}\left[\frac{x^3 g^{\prime\prime}(x)}{v_T}\left(f + \frac{T}{m_e v}\frac{\partial f}{\partial v}\right)\right]\Bigg\},
        \label{eq:C_ee0}
\end{split}
\end{equation}
where $x = v/v_T$, $g(x)$ is the function
\begin{equation}
        g(x) = \int \left|\mathbf{v} - \mathbf{v}^\prime\right|
        f_0\,\mathrm{d}^3v = n_0 v_T\left[\left(x +
        \frac{1}{2x}\right)\mathrm{erf}(x) + \frac{1}{\sqrt{\pi}}\mathrm{e}^{-x^2}\right],
\end{equation}
and
\begin{equation}
        n_0 = \eta \mathrm{e}^{\frac{e\phi}{T}}
\end{equation}
is the density of core electrons \citep{Helander2002}. We note that when $x$ is large, i.e. we consider collisions of an electron with an energy much larger than $T$ with the Maxwellian, then both $g^\prime(x)$ and $x^3 g^{\prime\prime}(x)$ are well-approximated by $n_0 v_T$. 

Similarly, assuming that collisions with ions are well-approximated by pitch-angle scattering, we have
\begin{equation}
        C_{e,ik}(f) = \frac{e^4 \ln\Lambda}{4\pi\varepsilon_0^2 m_e^2}\left[\frac{1}{v^3}Z_k^2 n_{ik}\frac{1}{2\sin\theta}\frac{\partial}{\partial \theta}\left(\sin\theta \frac{\partial f}{\partial \theta}\right)\right]  \label{eq:C_ik}
\end{equation}
for ion charge $Z_k$ and density $n_{ik}$. 

The collision operator is given by the divergence of the collisional flux $\mathbf{F}$ in velocity space:
\begin{equation}
    C(f,f_0) = \nabla_\mathbf{v}\cdot \mathbf{F},
\end{equation}
so it can always be expressed in the form
\begin{equation}
        C(f,f_0) = |J| \nabla_\mathbf{w}\cdot \tilde{\mathbf{F}},
\end{equation}
where $|J|$ is the Jacobian of the transformation between coordinates $\mathbf{v}$ and $\mathbf{w}$,
\begin{equation}
        J = \mathrm{det}\left(\frac{\partial w_i}{\partial v_j}\right),
\end{equation}
and $\tilde{\mathbf{F}}$ is the collisional flux in $\mathbf{w}$ phase-space. Noting that
\begin{equation}
    \mathrm{det}\left(\frac{\partial (\mathcal{E}_\parallel,\mathcal{E}_\perp,\varphi)}{\partial (\mathbf{v})} \right) = m_e^2 v_\parallel
\end{equation}
we seek to transform Eq.\,\eqref{eq:C_ee0} into the form
\begin{equation}
    C(f,f_0) = A v_\parallel \nabla_{(\mathcal{E}_\parallel, \mathcal{E}_\perp)}\cdot \tilde{\mathbf{F}}
\end{equation}
for some constant $A$. We find that
\begin{equation}
        \begin{split}
        C(f,f_0) = &\frac{e^4 \ln
        \Lambda}{4\pi\varepsilon_0^2 m_e}v_\parallel
                \nabla_{(\mathcal{E}_\parallel,\mathcal{E}_\perp)}\cdot \\ &\left[f_0\bigg(\mathbf{r}\mathbf{r}
        \mathcal{E}_\perp \frac{v_\parallel}{v^3}
        \frac{g^\prime(x)}{v_T}
                + T \frac{\mathbf{s}\mathbf{s}}{v^5
        v_\parallel} \frac{x^3 g^{\prime\prime(x)}}{v_T}\bigg)
        \nabla_{(\mathcal{E}_\parallel,\mathcal{E}_\perp)}\left(\frac{f}{f_0}\right)\right],
        \label{eq:C_ee02}
        \end{split}
\end{equation}
where
\begin{equation}
        \mathbf{r} = (1,-1),\quad \mathbf{s} =
        \left(v_\parallel^2,v_\perp^2\right)
\end{equation}
and $\mathbf{r}\mathbf{r}$, $\mathbf{s}\mathbf{s}$ represent dyadic products. Similarly,
\begin{equation}
        C_{e,ik}(f) = \frac{e^4 \ln
        \Lambda}{4\pi\varepsilon_0^2 m_e}v_\parallel
                \nabla_{(\mathcal{E}_\parallel,\mathcal{E}_\perp)}\cdot \left[f_0\left(\mathbf{r}\mathbf{r}
        \mathcal{E}_\perp \frac{v_\parallel}{v^3}
        Z_k^2 n_{ik}
                \right)\right].  \label{eq:C_ik_2}
\end{equation}
$\mathbf{r}\mathbf{r}$ is the tensor associated with pitch-angle scattering, altering parallel and perpendicular energies such that their sum is unchanged.
$\mathbf{s}\mathbf{s}$ is associated with energy exchange, which affects parallel and perpendicular
energies equally.

We must bounce-average the collision operators \eqref{eq:C_ee02} and \eqref{eq:C_ik_2}. In order to obtain a `useful' expression, we must be able to commute divergence in $(\mathcal{E}_\parallel,\mathcal{E}_\perp)$ and the orbit integral. The following Lemma is useful with regards to commuting the divergence and orbit integral: for a vector-valued function $\mathbf{F}$, we have
\begin{equation}
        \nabla_{(\mathcal{E}_\parallel, \mathcal{E}_\perp)}\cdot\oint \mathbf{F}\,\mathrm{d}z = \oint
        \nabla_{(\mathcal{E}_\parallel, \mathcal{E}_\perp)}\cdot \mathbf{F}\,\mathrm{d}z +
        4\nabla_{(\mathcal{E}_\parallel, \mathcal{E}_\perp)}z_c \cdot \mathbf{F}(z_c).
\end{equation}
In order for the divergence and orbit integral to commute, we must have
$\nabla_{(\mathcal{E}_\parallel,\mathcal{E}_\perp)}z_c \cdot \mathbf{F}(z_c) = 0$; either
$\mathbf{F}(z_c)$ or $\nabla_{(\mathcal{E}_\parallel,\mathcal{E}_\perp)}z_c$ vanishes, or $\nabla_{(\mathcal{E}_\parallel,\mathcal{E}_\perp)}z_c$ is
orthogonal to $\mathbf{F}(z_c)$.

We observe that the term associated with pitch-angle scattering is
proportional to $v_\parallel$, which (by definition) vanishes when $z = z_c$.
So, the divergence and orbit integral commute in the pitch-angle scattering
term.

As for the energy exchange term, we observe that 
\begin{equation}
        \nabla_{(\mathcal{E}_\parallel,\mathcal{E}_\perp)}z_c = \frac{\partial
        z_c}{\partial \mathcal{E}_\parallel}
        \left(\begin{matrix}
               1\\
               0
        \end{matrix}\right),
\end{equation}
and
\begin{equation}
        (\mathbf{s}\mathbf{s})(z=z_c) =
        \left(
        \begin{matrix}
                0 &0\\
                0 &v_\perp^4\\
        \end{matrix}
        \right),
\end{equation}
which means that with respect to this term,
\begin{equation}
        \nabla_{(\mathcal{E}_\parallel, \mathcal{E}_\perp)}z_c \cdot
        \mathbf{F}(z_c) \propto
        \left(
        \begin{matrix}
                1\\
                0
        \end{matrix}
        \right)\cdot
        \left[
        \left(
        \begin{matrix}
                0 &0\\
                0 &v_\perp^4
        \end{matrix}    
        \right)
        \nabla_{(\mathcal{E}_\parallel,
        \mathcal{E}_\perp)}\left(\frac{f}{f_0}\right)
        \right] = 0,
\end{equation}
so the orbit integral and divergence commute for the energy exchange term.
Since both $f$ and $f_0$ are independent of $z$, they and their
derivatives may be brought outside the orbit integral, giving the bounce-averaged QE collision operator
\begin{equation}
        \begin{split}
                \left<C_\mathrm{QE}(f)\right> = \frac{e^4 \ln
        \Lambda}{4\pi\varepsilon_0^2 m_e \tau}
                \nabla_{(\mathcal{E}_\parallel,\mathcal{E}_\perp)}\cdot\bigg[&f_0\bigg(\mathbf{r}\mathbf{r}
        \mathcal{E}_\perp \oint\frac{v_\parallel}{v^3}
        \left(\frac{g^\prime(x)}{v_T} + \sum_k Z_k^2
                n_{ik}\right)\,\mathrm{d}z \\
                &+ T \oint\frac{\mathbf{s}\mathbf{s}}{v^5
        v_\parallel} \frac{x^3 g^{\prime\prime(x)}}{v_T}\,\mathrm{d}z\bigg)
                \nabla_{(\mathcal{E}_\parallel,\mathcal{E}_\perp)}\left(\frac{f}{f_0}\right)\bigg].
        \label{eq:C_QE_ba}
        \end{split}
\end{equation}

The quasi-equilibrium equation is given by setting the above to zero:
\begin{equation}
        \begin{split}
                \nabla_{(\mathcal{E}_\parallel,\mathcal{E}_\perp)}\cdot\bigg[&f_0\bigg(\mathbf{r}\mathbf{r}
        \mathcal{E}_\perp \oint\frac{v_\parallel}{v^3}
        \left(\frac{g^\prime(x)}{v_T} + \sum_k Z_k^2
                n_{ik}\right)\,\mathrm{d}z \\
                &+ T \oint\frac{\mathbf{s}\mathbf{s}}{v^5
        v_\parallel} \frac{x^3 g^{\prime\prime(x)}}{v_T}\,\mathrm{d}z\bigg)
                \nabla_{(\mathcal{E}_\parallel,\mathcal{E}_\perp)}\left(\frac{f}{f_0}\right)\bigg]
                = 0,
        \label{eq:QE_ba}
        \end{split}
\end{equation}
which is in the form of an anisotropic steady-state diffusion problem in
$(\mathcal{E}_\parallel, \mathcal{E}_\perp)$ space:
\begin{equation}
        \nabla_{(\mathcal{E}_\parallel,\mathcal{E}_\perp)}\cdot
        \left[D_\mathrm{QE}
        \nabla_{(\mathcal{E}_\parallel,\mathcal{E}_\perp)}\left(\frac{f}{f_0}\right)\right] = 0 \label{eq:qe_4}
\end{equation}
for 
\begin{equation}
        D_\mathrm{QE} = D_\mathrm{QE,S} + D_\mathrm{QE,F},
\end{equation}
where the diffusion tensor associated with pitch-angle scattering is given by
\begin{equation}
        D_\mathrm{QE,S} = 
        f_0
                \mathcal{E}_\perp \oint\frac{v_\parallel}{v^3}
        \left(\frac{g^\prime(x)}{v_T} + \sum_k Z_k^2
                n_{ik}\right)\,\mathrm{d}z
        \left(
        \begin{matrix}
                1 &-1\\
                -1 &1
        \end{matrix}
        \right),
\end{equation}
and that associated with energy exchange is given by
\begin{equation}
        D_\mathrm{QE,F} = 
                f_0 T \oint\frac{1}{v^5
        v_\parallel}
        \left(
        \begin{matrix}
                v_\parallel^4 &v_\parallel^2 v_\perp^2\\
                v_\parallel^2 v_\perp^2 &v_\perp^4
        \end{matrix}
        \right)
        \frac{x^3 g^{\prime\prime}(x)}{v_T}\,\mathrm{d}z.
\end{equation}

Together with quasineutrality,
\begin{equation}
    \int f\,\mathrm{d}^3v = n_e = \sum_k Z_k n_{ik}, \label{eq:QN}
\end{equation}
Eq.\,\eqref{eq:qe_4} with the boundary conditions \eqref{eq:QE_BC_2},\eqref{eq:QE_BC_1} (noting that we write $f = f_\mathrm{II}$) provides a unique solution for $f$ and $\phi$ in terms of the parameters $\eta$ and $T$. However, these parameters are not know \textit{a priori}.

It should be noted that up to this point we have assumed that the potential is monotonically decreasing, which is the case when the density profile is monotonically decreasing. If this is not the case, i.e. the potential has more than one peak, then there are actually multiple trapped electron populations that must be treated independently. Some electrons can explore the region encompassed by only one peak, and others, still trapped in the potential as a whole, can explore more than one. This situation greatly complicates the kinetic problem and is of secondary importance in this paper as we are concerned with the expansion of a plasmoid with a potential that is initially single-peaked; we expect, and observe, as will be shown in Section \ref{sec:expansion_numerical}, the profile to remain single-peaked when the high temperature of the ambient plasma is accounted for. There is one exception to the inapplicability of the foregoing model to multiply-peaked electric potential wells: when $T = T_a$. In this case the solution to the QE problem is, anyway, the ambient Maxwellian $f_a$, which is correct even when multiple peaks are present. It will be seen that the cold-fluid model for ions does produce a multiply-peaked electric potential during later stages of the expansion, but at this point $T$ is of order $T_a$, so the solution to the QE problem is close to a Maxwellian and we expect the resulting expansion to be qualitatively correct.

\subsection{The no-net-flux condition}
Given some $\eta$ and $T$ we may solve the QE problem as specified in the previous subsection. However, most of the combinations of $\eta$ and $T$ are not physically meaningful, since they would not actually establish a steady-state. Quasineutrality requires that there is no net charge, but most combinations of $\eta$ and $T$ would cause a very large collisional flux of electrons into or out of the trapped region of phase-space, causing the plasmoid to `charge up', quickly resulting in the violation of global quasineutrality. Therefore a closer look at quasineutrality during the establishment of the QE state is required.

The global quasineutrality condition is given by
\begin{equation}
    N_t + N_p = \sum_k Z_k N_{ik}
\end{equation}
for $N_t$ the line-integrated density of trapped electrons, $N_p$ the line-integrated density of passing electrons, and $N_{ik}$ the line-integrated density of ion species $k$. Formally, the magnetic field line we consider is infinite. However, the entire plasmoid structure is localised, with the possible exception of the plasmoid density approaching zero asymptotically (if we use, for example, the self-similar ion density profile from Ref.\,\cite{Aleynikov2019}). So, rather than the whole field line, we consider the global quasineutrality condition on some interval $z \in [-L_S/2,L_S/2]$ for some $L_S$ much larger than the plasmoid; large enough that the plasmoid density at the endpoints is negligible compared to the ambient density, and the electric potential at the endpoints is negligible compared to $T$, $T_a$, or $e\phi_m$.

In order to maintain global quasineutrality we require
\begin{equation}
    \frac{\mathrm{d} N_t}{\mathrm{d} t} + \frac{\mathrm{d} N_p}{\mathrm{d} t} = \sum_k Z_k \frac{\mathrm{d} N_{ik}}{\mathrm{d} t}. \label{eq:global_QN_time_deriv}
\end{equation}
Since the density is constant far the plasmoid, the terms in the above cease to change as $L_S \rightarrow \infty$; we may take $L_S$ arbitrarily large as we never directly evaluate $N_p$ or $N_{ik}$.

Using the results from Appendix A, we see that the time derivative of the line-integrated density of trapped electrons is given by
\begin{equation}
    \frac{\mathrm{d} N_t}{\mathrm{d} t} = \frac{2\pi}{m_e^2} \frac{\mathrm{d}}{\mathrm{d} t} \int^\infty_0 \int^{J_m}_0 f\,\mathrm{d}J\,\mathrm{d}\mathcal{E}_\perp,
\end{equation}
where $J_m = J(\mathcal{E}_\parallel = 0)$ is the maximum value of the second adiabatic invariant for a trapped electron, and, knowing that $f = f_a$ in the $\mathcal{E}_\parallel > 0$ region, the time derivative of the line-integrated density of passing electrons is given by
\begin{equation}
    \frac{\mathrm{d} N_p}{\mathrm{d} t} = \frac{\mathrm{d}}{\mathrm{d} t}\int^\frac{L_S}{2}_{-\frac{L_S}{2}} n_a \mathrm{e}^{\frac{e\phi}{T_a}}\mathrm{erfc}\left(\sqrt{\frac{e\phi}{T_a}}\right)\,\mathrm{d}z.
\end{equation}
The full kinetic equation \eqref{eq:kinetic_2} can be bounce-averaged and the left hand side changed to the independent variables $(J,\mathcal{E}_\perp,t)$ to yield
\begin{equation}
    \frac{\partial f}{\partial t}\bigg|_J = \left<C(f,f_\mathrm{I})\right> + \sum_k \left<C_{e,ik}(f)\right> + \left<C(f,f_\mathrm{II} + f_\mathrm{III})\right>,
\end{equation}
which, along with the approximation $C(f,f_\mathrm{I}) \approx C(f,f_0)$, gives
\begin{equation}
\begin{split}
    \frac{\mathrm{d} N_t}{\mathrm{d} t} = &\frac{2\pi}{m_e^2} \int^\infty_0 \int^{J_m}_0 \left<C_\mathrm{QE}(f)\right>\,\mathrm{d}J\,\mathrm{d}\mathcal{E}_\perp + \\ &\frac{2\pi}{m_e^2} \int^\infty_0 \int^{J_m}_0 \left<C(f,f_\mathrm{II}+f_\mathrm{III})\right>\,\mathrm{d}J\,\mathrm{d}\mathcal{E}_\perp + \\ &\frac{2\pi}{m_e^2} \int^\infty_0 \frac{\partial J_m}{\partial t} f_a(\mathcal{E}_\perp)\,\mathcal{E}_\perp,
\end{split}
\end{equation}
where we have used the fact that the distribution function is continuous: $f(J = J_m) = f(\mathcal{E}_\parallel = 0) = f_a(\mathcal{E}_\parallel = 0)$. We then see that Eq.\,\eqref{eq:global_QN_time_deriv} becomes
\begin{equation}
\begin{split}
        &\frac{2\pi}{m_e^2} \int^\infty_0 \int^{J_m}_0 \left<C_\mathrm{QE}(f)\right>\,\mathrm{d}J\,\mathrm{d}\mathcal{E}_\perp + \\ &\frac{2\pi}{m_e^2} \int^\infty_0 \int^{J_m}_0 \left<C(f,f_\mathrm{II}+f_\mathrm{III})\right>\,\mathrm{d}J\,\mathrm{d}\mathcal{E}_\perp + \\ & \int^{\frac{L_S}{2}}_{-\frac{L_S}{2}}n_a \frac{e}{T_a}\frac{\partial \phi}{\partial t}\mathrm{e}^{\frac{e\phi}{T_a}}\mathrm{erfc}\left(\sqrt{\frac{e\phi}{T_a}}\right)\,\mathrm{d}z
        = \sum_k Z_k \frac{\mathrm{d} N_{ik}}{\mathrm{d} t}, \label{eq:no_net_flux_intermediate}
\end{split}
\end{equation}
where the term associated with $\partial J_m/\partial t$ has cancelled out between $\mathrm{d}N_t/\mathrm{d}t$ and $\mathrm{d}N_p/\mathrm{d}t$.

The first term on the left hand side is associated with fluxes due to collisions with the plasmoid; we may write
\begin{equation}
    \frac{2\pi}{m_e^2} \int^\infty_0 \int^{J_m}_0 \left<C_\mathrm{QE}(f)\right>\,\mathrm{d}J\,\mathrm{d}\mathcal{E}_\perp \sim \nu_p(v_{T_a})N_p.
\end{equation}
The second term on the left hand side is due to heating; we may write
\begin{equation}
    \frac{2\pi}{m_e^2} \int^\infty_0 \int^{J_m}_0 \left<C(f,f_\mathrm{II}+f_\mathrm{III})\right>\,\mathrm{d}J\,\mathrm{d}\mathcal{E}_\perp \sim \nu_h \mathrm{e}^{-\nu_h t} N_p.
\end{equation}
The term on the right hand side is due to the plasma at infinity acting as a source or sink of ions.

The third term on the left hand side is due to the constant replenishment of the passing distribution by plasma at infinity, leading to it always having the form $f_a$. We can estimate this term by noting that $\exp(e\phi/T_a)\mathrm{erfc}(\sqrt{e\phi/T_a}) \leq 1$ for $\phi \ge 0$ and using the approximation
\begin{equation}
    \int^{\frac{L_S}{2}}_{-\frac{L_S}{2}} \frac{e\phi}{T_a}\,\mathrm{d}z \sim L_p \frac{e\phi_m}{T_a} \sim L_p. 
\end{equation}
Writing $L_p = N_p/n_m$ and using Eq.\,\eqref{eq:n_m_Aleynikov_modified} then yields
\begin{equation}
    \int^{\frac{L_S}{2}}_{-\frac{L_S}{2}} n_a \frac{e}{T_a}\frac{\partial \phi}{\partial t}\mathrm{e}^{\frac{e\phi}{T_a}}\mathrm{erfc}\left(\sqrt{\frac{e\phi}{T_a}}\right)\,\mathrm{d}z \sim \nu_h N_p \frac{\mu \sqrt{\frac{3m_i}{m_e} \nu_h t}}{\left((\nu_h t)^\frac{3}{2} + \frac{2}{\sqrt{3}}\sqrt{\frac{m_i}{m_e}}\mu\right)^2}. \label{eq:global_QN_third}
\end{equation}
With the plasma parameters used in the Introduction, the above is at most of order $\nu_h N_p$, and decreases as time advances; it is of the same order as the heating term (the second on the left hand side of Eq.\,\eqref{eq:no_net_flux_intermediate}).

The change in $N_{ik}$ depends upon the (yet unchosen) model for the ions. Of course, the system for plasmoid ions is necessarily conservative (plasmoid ions are localised), so the line-integrated plasmoid ion density is constant. If the system also conserves ambient ions, i.e. the ambient plasma cannot act as a source of ions, then the line-integrated ambient ion densities are constant. On the other hand, if the plasma can act as a source for the ambient ions then the terms associated with the change in their line-integrated densities is at most on the same order as Eq.\,\eqref{eq:global_QN_third}. This is because an ambient ion density $n_{ik,a}$ is at most of order $n_a/Z_k$ and its change is associated purely with the change in the electric potential.

Therefore the most significant term in Eq.\,\eqref{eq:no_net_flux_intermediate} is the first on the left hand side, which is associated with frequency with which an electron collides with the plasmoid; the other terms are associated with the heating frequency. Hence Eq.\,\eqref{eq:no_net_flux_intermediate} is well-approximated with simply a vanishing flux associated with collisions with the plasmoid, which can be expressed as
\begin{equation}
    \frac{e^4 \ln \Lambda}{2\varepsilon_0^2 m_e^3}
        \int^\infty_0 \left[D_\mathrm{QE}
        \nabla_{(\mathcal{E}_\parallel,\mathcal{E}_\perp)}\left(\frac{f}{f_0}\right)\right]\bigg|_{\mathcal{E}_\parallel = 0}\cdot
        \left(\begin{matrix}
            1 \\
            0
        \end{matrix}\right)
        \,\mathrm{d}\mathcal{E}_\perp = 0.\label{eq:no_net_flux}
\end{equation}
Hence, to maintain global quasineutrality, there can (approximately) be no net collisional flux due to collisions with the plasmoid into the trapped region of phase-space; we call the above the \textit{no-net-flux} condition. Intuitively this makes sense; a steady-state due entirely to collisional fluxes cannot exist if a consequence of the flux is an immediate violation of quasineutrality.

Since we had two free parameters, $\eta$ and $T$, the no-net-flux condition fixes one in terms of the other. We choose to keep $T$ as the free parameter. From Eq.\,\eqref{eq:qe_4} it is clear that although the QE problem implies a steady-state, the collisional fluxes themselves do not vanish. This is the sense in which QE is a dynamical steady-state rather than the static steady-state characteristic of a thermal equilibrium.

\subsection{Numerical solution to the QE problem} \label{sec:qe_numerical}
The QE problem \eqref{eq:qe_4} with boundary conditions \eqref{eq:QE_BC_2}, \eqref{eq:QE_BC_1} (noting that we write $f = f_\mathrm{II}$ in region II), a self-consistent potential given by quasineutrality \eqref{eq:QN}, and the zero-net-flux condition \eqref{eq:no_net_flux} was solved numerically. The plasma was assumed to be hydrogenic: there was a single species of singly-charged ion ($Z = 1$) with the proton mass ($m_i$ = $m_p$), following the profile
\begin{equation}
    n_i = n_a + N_{ic}\frac{1}{L_p\sqrt{\pi}}\mathrm{e}^{-\left(\frac{z}{L_p}\right)^2},
\end{equation}
where $L_p = 2.8\,\mathrm{m}$, $N_{ic} = 10^{22}\,\mathrm{m}^{-2}$, $T = 615\,\mathrm{eV}$, $n_a = 5\times10^{19}\,\mathrm{m}^{-3}$, and $T_a = 5\,\mathrm{keV}$. The Gaussian term in the above is consistent with the profile in \cite{Aleynikov2019} at $t = 20\,\mathrm{\mu s}$ given these parameters.

\begin{figure}
    \centering
    \hspace{0.5cm}
    \includegraphics[width=\textwidth]{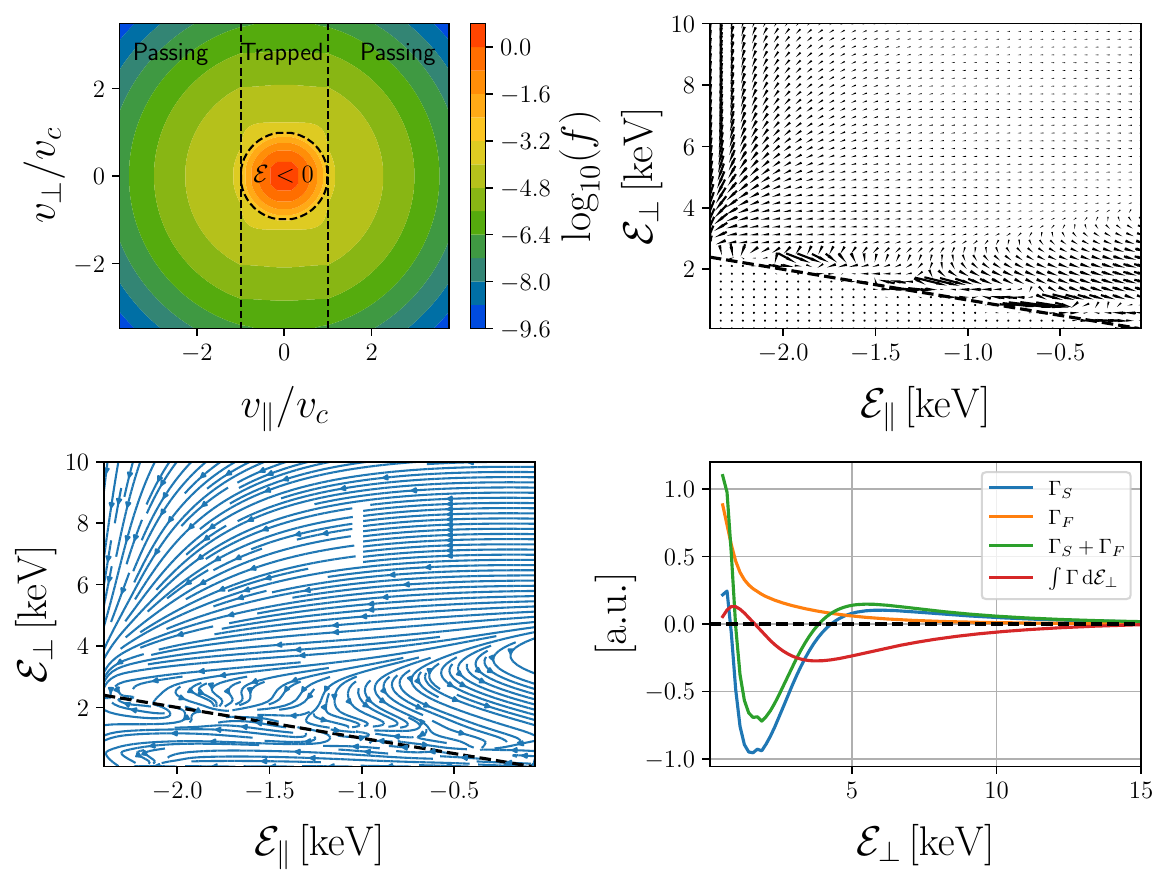}
    \caption{Top left: Numerical distribution function at $z = 0$ in SI units. Top right: effective phase-space flow velocity $\mathbf{u}^*$ (Eq.\,\eqref{eq:ustar}). Bottom left: phase-space trajectories of electrons (i.e. the streamlines of $\mathbf{u}^*$). Bottom right: collisional flux into the trapped region (Eqs.\,\eqref{eq:fluxS},\eqref{eq:fluxF}). $v_c = \sqrt{2\phi_m/m_e}$ is the parallel escape velocity at $z = 0$.}
    \label{fig:1e22-20mus-f}
\end{figure}

Figure \ref{fig:1e22-20mus-f} shows the properties of the electron distribution function. The top left plot shows the distribution function in velocity space at $z = 0$. $\mathcal{E} = 0$ is indicated by the dashed circle and the trapped-passing separatrix by the vertical dashed lines. The isotropic passing distribution is clearly visible as concentric circles, as is the very isotropic core ($\mathcal{E} < 0$). Significant flattening of the distribution in the high-energy part region II is observed. The bottom left plot shows the phase-space trajectories of electrons in $(\mathcal{E}_\parallel,\mathcal{E}_\perp)$ space, clearly indicating flow into and out of the trapped region. The dashed line indicates $\mathcal{E} = 0$. The right boundary is the trapped-passing separatrix. The top right plot shows the effective phase-space flow velocity $\mathbf{u}^*$, defined via
\begin{equation}
    \nabla_{(\mathcal{E}_\parallel,\mathcal{E}_\perp)} \cdot \left(\mathbf{u}^*f\right) = - \left<C_\mathrm{QE}(f)\right>,
\end{equation}
i.e.
\begin{equation}
    \mathbf{u}^* \propto -\frac{D_\mathrm{QE} \nabla_{(\mathcal{E}_\parallel,\mathcal{E}_\perp)} \left(\frac{f}{f_0}\right)}{f}. \label{eq:ustar}
\end{equation}
Electron phase-space flow for $\mathcal{E} < 0$ is very weak, since this region is highly isotropised and essentially conforms to a Maxwellian, which exhibits no collisional phase-space flux. 

The bottom right plot shows the flux through the trapped-passing separatrix, where
\begin{equation}
    \Gamma_S = -\frac{e^4 \ln \Lambda}{2\varepsilon_0^2 m_e^3}\left[D_\mathrm{QE, S}
        \nabla_{(\mathcal{E}_\parallel,\mathcal{E}_\perp)}\left(\frac{f}{f_0}\right)\right]\bigg|_{\mathcal{E}_\parallel = 0}\cdot
        \left(\begin{matrix}
            1 \\
            0
        \end{matrix}\right)\label{eq:fluxS}
\end{equation}
\begin{equation}
    \Gamma_F = -\frac{e^4 \ln \Lambda}{2\varepsilon_0^2 m_e^3}\left[D_\mathrm{QE, F}
        \nabla_{(\mathcal{E}_\parallel,\mathcal{E}_\perp)}\left(\frac{f}{f_0}\right)\right]\bigg|_{\mathcal{E}_\parallel = 0}\cdot
        \left(\begin{matrix}
            1 \\
            0
        \end{matrix}\right), \label{eq:fluxF}
\end{equation}
and $\Gamma = \Gamma_S + \Gamma_F$. Collisions with the cold Maxwellian always produces an inflow of electrons in region II due to friction (see $\Gamma_F$ in Fig.\,\ref{fig:1e22-20mus-f}). Pitch-angle scattering may only cause a flow along lines of constant $\mathcal{E}$, and is seen to eject electrons at low perpendicular energies and causes an inflow at higher energies. The net flux through the separatrix vanishes due to the no-net-flux condition.

\FloatBarrier

\subsection{Analytical solution to the QE problem} \label{sec:qe_approx}
The main difficulty in the QE problem is the presence of bounce integrals, which are difficult to evaluate analytically in a self-consistent potential. In a square well, however, owing to the constancy of the electric potential and plasmoid density, the bounce average operator is the identity, significantly simplifying matters. Since the phase-space domain of the QE problem depends only upon the potential height $\phi_m$, the distribution function obtained as a solution to the QE problem in a square-well potential of height $\phi_m$ may be used as an approximation of the solution to the QE problem in a self-consistent potential.

Assuming that there is a single ion species of charge $Z$, that hot trapped electrons have energies much larger than $v_T$, and that the plasmoid density greatly exceeds the ambient density (hence quasineutrality is approximately given by $n_0 = Z n_i$), from Eqs.\,\eqref{eq:C_ee0},\eqref{eq:C_ik} we see that the bounce-averaged QE collision operator in a square well is given by
\begin{equation}
   \left<C_\mathrm{QE}(f)\right> = \frac{n_0 e^4 \ln\Lambda}{4\pi \varepsilon_0^2 m_e^2}\left[\frac{1 + Z}{v^3}\frac{1}{2\sin\theta}\frac{\partial}{\partial \theta}\left(\sin\theta \frac{\partial f}{\partial \theta}\right) + \frac{1}{v^2}\frac{\partial}{\partial v}\left(f + \frac{T}{m_e v}\frac{\partial f}{\partial v}\right)\right]. \label{eq:qe_squarewell}
\end{equation}
We consider $\left<C_\mathrm{QE}(f)\right> = 0$ separately in the $\mathcal{E} < 0$ (i.e. $v < v_c = \sqrt{2e\phi_m/m_e}$ in a square well) and $\mathcal{E} > 0$ ($v > v_c$) regions of phase-space, using the most convenient coordinates in each case. It is convenient to write $f = f_0 + f_1$ and solve for $f_1$; in both regions we neglect $v$-diffusion for $f_1$ (i.e. the term proportional to $T/(m_e v)$ in the above), leaving only $v$-friction.

In the $\mathcal{E} > 0$ region we use the variables $(v,v_\parallel = v\cos\theta)$, meaning we must solve $\mathcal{D}(f_1) = 0$ for
\begin{equation}
    \mathcal{D}(f_1) = \frac{1 + Z}{2}\frac{\partial}{\partial v_\parallel}\left[(v^2 - v_\parallel^2)\frac{\partial f_1}{\partial v_\parallel}\right] + v \frac{\partial f_1}{\partial v} + v_\parallel \frac{\partial f_1}{\partial v_\parallel}.
\end{equation}
It is convenient to consider the limit where $v \gg v_\parallel$, which represents the correct limit in the majority of $\mathcal{E} > 0$, $\mathcal{E}_\parallel < 0$ phase-space:
\begin{equation}
    \mathcal{D}(f_1) = \frac{1 + Z}{2}v^2\frac{\partial^2 f_1}{\partial v_\parallel^2} + v \frac{\partial f_1}{\partial v},
\end{equation}
which has superposable solutions (that are even in $v_\parallel$) provided by the separation of variables:
\begin{equation}
    f_k = C_k \mathrm{e}^{-\frac{1}{2}\lambda_k v^2}\cosh\left(v_\parallel\sqrt{\frac{2}{1+Z}\lambda_k}\right)
\end{equation}
for constants $\left\{C_k\right\}$ and $\left\{\lambda_k\right\}$. The boundary condition \eqref{eq:QE_BC_1} is satisfied by the sum of two solutions:
\begin{equation}
    f_1 =  f_a(\mathcal{E}) \frac{\cosh\left(\frac{v_\parallel}{v_{T_a}}\sqrt{\frac{4}{1+Z}}\right)}{\cosh\left(\frac{v_c}{v_{T_a}}\sqrt{\frac{4}{1+Z}}\right)} - f_0(\mathcal{E}) \frac{\cosh\left(\frac{v_\parallel}{v_{T}}\sqrt{\frac{4}{1+Z}}\right)}{\cosh\left(\frac{v_c}{v_{T}}\sqrt{\frac{4}{1+Z}}\right)}. \label{eq:f1_energypassing}
\end{equation}

In the $\mathcal{E} < 0$ region we use the variables variables $(v,\xi = \cos\theta)$. So, we must solve $\mathcal{G}(f_1) = 0$ where
\begin{equation}
    \mathcal{G}(f_1) = \frac{1 + Z}{2}\frac{\partial}{\partial \xi}\left[\left(1 - \xi^2\right)\frac{\partial f_1}{\partial \xi}\right] + v \frac{\partial f_1}{\partial v}.
\end{equation}
We note that Legendre polynomials are the eigenfunctions of the operator in $\xi$, so we write $f$ in this basis:
\begin{equation}
    f_1 = \sum_{n=0}^\infty a_n(v) P_n(\xi), \label{eq:f_approx_energytrapped}
\end{equation}
which gives the following equations for $\{a_n(v)\}$:
\begin{equation}
    v \frac{\partial a_n}{\partial v} - \frac{1 + Z}{2}n(n+1) a_n = 0
\end{equation}
with the solutions
\begin{equation}
    a_n = c_n v^{\frac{1 + Z}{2}n(n+1)}
\end{equation}
for $\{c_n\}$ constants. The continuity of the distribution function at $v = v_c$ provides the expressions for $c_n$:
\begin{equation}
\begin{split}
    c_n = \frac{2n + 1}{2}v_c^{-\frac{1+Z}{2}n(n+1)}\Bigg[\frac{f_a(\mathcal{E}=0)}{\mathrm{cosh}\left(\frac{v_c}{v_{T_a}}\sqrt{\frac{4}{1+Z}}\right)}\int^1_{-1}\mathrm{cosh}\left(\frac{v_c}{v_{T_a}}\sqrt{\frac{4}{1+Z}}\xi\right)P_n(\xi)\,\mathrm{d}\xi - \\
    \frac{f_0(\mathcal{E}=0)}{\mathrm{cosh}\left(\frac{v_c}{v_{T}}\sqrt{\frac{4}{1+Z}}\right)}\int^1_{-1}\mathrm{cosh}\left(\frac{v_c}{v_{T}}\sqrt{\frac{4}{1+Z}}\xi\right)P_n(\xi)\,\mathrm{d}\xi\Bigg].
\end{split}
\end{equation}
To summarise, the analytical solution to the QE problem in a square well is given by Eq.\,\eqref{eq:f1_energypassing} in the $\mathcal{E} > 0$, $\mathcal{E}_\parallel < 0$ region, Eq.\,\eqref{eq:f_approx_energytrapped} in the $\mathcal{E} \le 0$ region, and $f = f_a$ in the $\mathcal{E}_\parallel \ge 0$ region. The phase-space domain of the QE problem is the same given any potential, so the substitution of $\sqrt{(2/m_e)(\mathcal{E} + e\phi_m)}$ for $v$ and $\sqrt{(2/m_e)(\mathcal{E}_\parallel + e\phi_m)}$ for $v_\parallel$ yields an (approximate) analytical solution to the QE problem valid in a self-consistent potential. We refer to this analytical solution in figures as $f_\mathrm{an}$.

\begin{figure}
    \centering
    \hspace{0.5cm}
    \includegraphics[width=\textwidth]{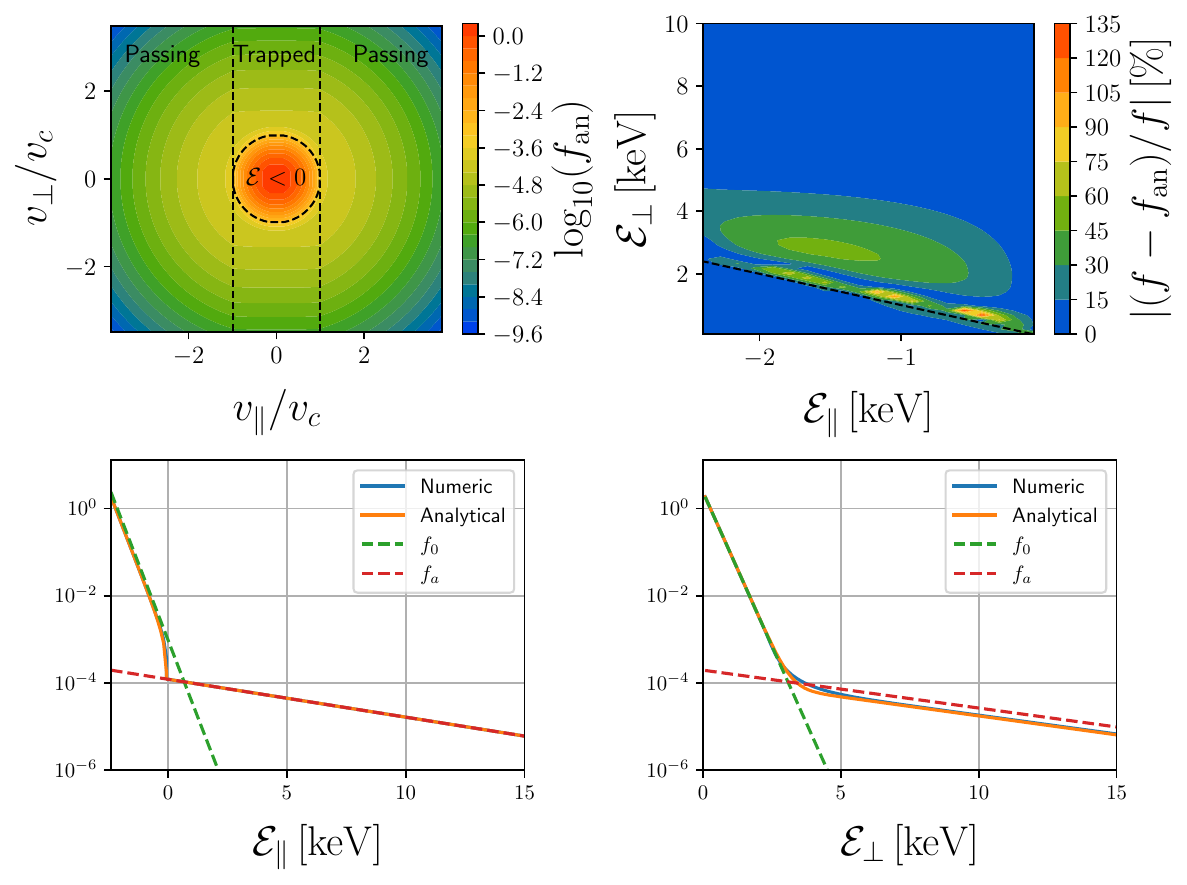}
    \caption{Top left: analytical electron distribution $f_\mathrm{an}$ in SI units. Top right: percentage difference between the numerical $f$ (see Fig.\,\ref{fig:1e22-20mus-f}) and analytical $f_\mathrm{an}$. Bottom left: distributions at $\mathcal{E}_\perp = 0$. Bottom right: distributions at $\mathcal{E}_\parallel = -e\phi_m$. $v_c = \sqrt{2\phi_m/m_e}$ is the parallel escape velocity at $z = 0$.}
    \label{fig:1e22-20mus-fanalytical}
\end{figure}

Figure \ref{fig:1e22-20mus-fanalytical} shows the analytical distribution function for the same parameters as those used to produce Fig.\,\ref{fig:1e22-20mus-f}. The top left plot shows the distribution in velocity space at $z = 0$. The top right plot shows the percentage difference from the numerical solution given in Fig.\,\ref{fig:1e22-20mus-f}. The bottom left plot shows the distribution function at $\mathcal{E}_\perp = 0$. The bottom right plot shows the distribution function at $\mathcal{E}_\parallel = -e\phi_m$. The qualitative behaviour of the distribution is captured well, in particular the `flattening' of the distribution function in the high-energy part of region II. The observation that in here the contours of the distribution function are horizontal lines in $(v_\parallel,v_\perp)$ can be explained by the fact that for the analytical QE distribution function
\begin{equation}
    \frac{\partial}{\partial v_\parallel}f\left(v_\parallel, v_\perp\right) \propto \frac{2}{v_{T_a}^2} v_\parallel\left(\frac{2}{1 + Z} - 1\right) + O(v_\parallel^3),
\end{equation}
which vanishes to lowest order if $Z = 1$.

We note that the simplification made by neglecting the $v$-diffusion term for $f_1$ reduces a formerly second-order problem in $v$ to a first-order problem. Solving the QE problem in the $\mathcal{E} > 0$ part of region II with the continuity boundary condition then fixes the value of $f$ at $\mathcal{E} = 0$. Similarly, in the $\mathcal{E} < 0$ part of region II we only have the opportunity to apply the continuity boundary condition at $\mathcal{E} = 0$, but not at $\mathcal{E} = \mathcal{E}_\mathrm{I/II}$. As a consequence, as $\mathcal{E} \rightarrow \mathcal{E}_\mathrm{I/II}$, $f_1 \rightarrow c_0$, which does violate this continuity boundary condition. On the other hand, $c_0$ is several orders of magnitude smaller than $f_0$ in region $I$, so the discontinuity is small. This is purely an artefact of the approximations made to obtain the analytical solution; no such behaviour is seen in the numerical solution.

\subsection{Expressions for $\eta$ and $\phi$ in terms of $T$}
We have found an analytical solution to the QE problem in terms of the lowest-order distribution function $f_0$, which is uniquely determined by the parameters $\eta$ and $T$. The electric potential $\phi$ is as yet unknown, but must be such that quasineutrality \eqref{eq:QN} is satisfied. Additionally, the no-net-flux condition \eqref{eq:no_net_flux} must be satisfied. Therefore, we may reduce the number of unknowns in the system from three ($\eta$, $T$, and $\phi$) to one (which we choose to be $T$) by imposing quasineutrality and no-net-flux.

The analytical solution to QE, however, does not capture the `ejection structure' near $\mathcal{E}_\parallel = \mathcal{E}_\perp = 0$ characterising the flux of electrons out of trapped phase-space, so the no-net-flux condition may not be used directly. Instead, analysis of the flow of electrons in phase-space allows us to formulate a condition that is approximately equivalent to Eq.\,\eqref{eq:no_net_flux}. We note that, owing to the flattening of the distribution function high-energy part of region II, pitch-angle scattering acts to scatter newly-trapped electrons to lower parallel velocity while leaving their energy unchanged. Simultaneously the scattered electrons lose energy via friction. This can be seen in the streamline plot of Fig.\,\ref{fig:1e22-20mus-f}. The net effect is that a large fraction of electrons entering the trapped region of phase-space are eventually drawn into the $\mathcal{E}<0$ region. In order for there to be no net flux into the trapped region, the same number of electrons entering the $\mathcal{E} < 0$ region must escape from it, eventually being scattered and ejected from trapped phase-space (forming the ejection structure).

Therefore, an approximation to the condition \eqref{eq:no_net_flux}, which states that there is no net flux into the trapped region of phase-space, is that there is no net flux into the $\mathcal{E} < 0$ region of phase-space. Since the flux into and out of this region is characterised by integrals on the $\mathcal{E} = 0$ line it is not necessary to have a description of the ejection structure, whereas directly applying Eq.\,\eqref{eq:no_net_flux} requires us to integrate along $\mathcal{E}_\parallel = 0$, where the ejection structure strongly affects the flux. Since the analytical solution to QE potentially has a discontinuous derivative at $\mathcal{E} = 0$, the flux into the $\mathcal{E} < 0$ region has contributions from $\mathcal{E} \rightarrow 0^+$ and $\mathcal{E} \rightarrow 0^-$. We apply this approximation of the no-net-flux condition to the analytical solution to QE in a square well.

Since the analytical solution to the QE problem is derived from that posed in a square well, we will calculate the flux in terms of the variables $(v,\xi,\varphi)$. Pitch-angle scattering may not change the energy of electrons, so the pitch-angle scattering terms of the collision operator cannot contribute to the flux across $\mathcal{E} = 0$.  In a square well, the number of electrons entering the $\mathcal{E} < 0$ region of phase-space is equal to the number of electrons entering the $v < v_c$ region of phase space. From Eq.\,\eqref{eq:qe_squarewell} we see that the collisional flux in the $\hat{\mathbf{v}}$ direction at $v = v_c$ (assuming that $v_c \gg v_T$) is given by
\begin{equation}
    F = -\frac{n_0 e^4 \ln\Lambda}{4\pi\varepsilon_0^2 m_e^2} \frac{1}{v_c^2}\left(f + \frac{T}{m_e v}\frac{\partial f}{\partial v}\right)\Bigg|_{\mathcal{E} = 0}.
    \label{eq:F}
\end{equation}
The phase-space area element on the $v = v_c$ sphere is given by $v_c^2 \mathrm{d}\xi\,\mathrm{d}\varphi$, so the net flux into the $v < v_c$ region is given by
\begin{equation}
    G = \int^{2\pi}_0 \int^1_{-1} v_c^2\left(F\big|_{\mathcal{E} = 0^+} + F\big|_{\mathcal{E} = 0^-}\right)\,\mathrm{d}\xi\,\mathrm{d}\varphi.
\end{equation}
Setting the above to zero will provide the relation $\eta$, $T$, and $\phi$ required for no net flux into the $\mathcal{E}<0$ region. It will be seen that the expression obtained from $G = 0$ in the limit $Z \rightarrow \infty$ is quite accurate, even for $Z = 1$. In this limit the analytical solution is given by $f = f_a$ for $\mathcal{E} > 0$ and $f = c_0 + f_0$ for $\mathcal{E} < 0$, where $c_0 = f_a(\mathcal{E} = 0) - f_0(\mathcal{E} = 0)$. Then, $G = 0$ yields the relation
\begin{equation}
    \eta = n_a \left(\frac{T}{T_a}\right)^\frac{3}{2}\left(2 - \frac{T}{T_a}\right). \label{eq:eta_approx}
\end{equation}

Now, quasineutrality allows us to express both $\eta$ and $\phi$ in terms of $T$ and the ion profile. In the variables $(\mathcal{E}_\parallel, \mathcal{E}_\perp,z,t)$ the electron distribution is independent of $z$, so the electron density is a function of $\phi$, time, and the parameter $T$ ($\eta$ being already expressed in terms of $T$ via the above). Therefore the quasineutrality condition \eqref{eq:QN} demands that
\begin{equation}
    n_e(\phi,t;T) = \sum_k Z_k n_{ik}, \label{eq:QN_2}
\end{equation}
which gives an implicit expression for $\phi$ at every point, given some $T$ and some ion density profiles $\{n_{ik}\}$. We can immediately obtain an expression for the potential height when the plasmoid density is high; in this case the electron density at $z = 0$ is well-approximated by
\begin{equation}
    n_e(z = 0) = \eta \mathrm{e}^\frac{e\phi_m}{T},
\end{equation}
so
\begin{equation}
    e\phi_m = T\left[\ln\left(\frac{n_e(z=0)}{n_a}\right) + \frac{3}{2}\ln\left(\frac{T_a}{T}\right) - \ln\left(2 - \frac{T}{T_a}\right)\right]. \label{eq:phim_estimate}
\end{equation}
This is a modification of the Boltzmann relation for a plasma in thermal equilibrium, the extra contributions accounting for the fact that in the  quasi-equilibrium state the distribution function is not necessarily Maxwellian. As expected, the above reduces to the Boltzmann relation for $T = T_a$ (when the distribution becomes Maxwellian), agreeing with the rough estimate \eqref{eq:Boltzmann} which was used to justify the ordering leading to QE. As noted earlier, $e\phi_m$ given by \eqref{eq:phim_estimate} exceeds that given by the Boltzmann relation when $T < T_a$.

The height of the potential in the numerical solution to the QE problem (Fig.\,\ref{fig:1e22-20mus-f}), which is calculated entirely self-consistently, and leads to no net flux into the trapped region, is within 3\% of the estimate \eqref{eq:phim_estimate}. This is a remarkable agreement considering that the estimate was derived in the limit $Z \rightarrow \infty$, but the numerical solution is with $Z = 1$. The estimate also correctly predicts the height of the self-consistent potential for the solution to the time-dependent kinetic problem with isotropic electrons given in \cite{Arnold2023}, which is strong evidence that the QE state was established there too.

\FloatBarrier

\section{Plasmoid expansion} \label{sec:expansion}
\subsection{Treatment of ions}
Two different models for the ions are considered: a collisionless (Vlasov) system with hot ambient ions but cold plasmoid ions, and a cold-fluid expansion. We restrict our attention to a single ion species of charge $Z$. The collisionless and fluid models are in opposite regimes of collisionality, which will provide the broadest range of qualitative results for the plasmoid expansion. The collisionless system is, however, the most physically accurate model for the plasmoid expansion since ambient ions have a long mean-free-path relative to the plasmoid size. In fact, the ratio of plasmoid size to the ambient ion mean free path is the same as ambient electrons, expressed by the opacity \eqref{eq:opacity}. The collisionless system will still model cold plasmoid ions accurately since they will be initialised with zero velocity spread.

For the collisionless ion expansion we solve the Vlasov equation for ions:
\begin{equation}
    \frac{\partial f_i}{\partial t} + v_\parallel \frac{\partial f}{\partial z} - \frac{Z e}{m_i}\frac{\partial \phi}{\partial z}\frac{\partial f_i}{\partial v_\parallel} = 0. \label{eq:ion_vlasov}
\end{equation}

For the cold-fluid expansion we solve the equations
\begin{equation}
    \frac{\partial n_i}{\partial t} + \frac{\partial}{\partial z}(n_i u_i) = 0, \label{eq:ion_continuity}
\end{equation}
\begin{equation}
    \frac{\partial u_i}{\partial t} + u_i \frac{\partial u_i}{\partial z} + \frac{Z e}{m_i}\frac{\partial \phi}{\partial z} = 0. \label{eq:ion_velocity}
\end{equation}

\subsection{Treatment of electrons on expansion timescale}
Equations \eqref{eq:kinetic_region_I_higher} and \eqref{eq:kinetic_region_II_ba_higher} describe the electron dynamics on the expansion and heating timescales. We note that because they take exactly the same form, Eq.\,\eqref{eq:kinetic_region_II_ba_higher} and the bounce average of Eq.\,\eqref{eq:kinetic_region_I_higher} may be combined to yield
\begin{equation}
    \frac{\partial f}{\partial t} - \frac{1}{\tau}\frac{\partial J}{\partial t}\frac{\partial f}{\partial \mathcal{E}_\parallel} = \left<C(f,f_\mathrm{II} + f_\mathrm{III})\right> \label{eq:kinetic_expansion}
\end{equation}
to describe the both regions I and II (writing the electron distribution in both regions as $f$). This equation describes the long-term expansion of the plasmoid, which is driven by the heating term on the right hand side.

By solving the QE problem with the restrictions of quasineutrality and the no-net-flux condition, the distribution is known up to the parameter $T$; the evolution of the electron distribution function is completely characterised by how $T$ changes in time. In analogy with the Braginskii equations we obtain an evolution equation for $T$ by taking moments of the kinetic equation over phase-space. When we take moments the quasi-equilibrium distribution serves the same role as the near-Maxwellian distribution in the Braginskii equations.

In contrast to a \textit{local} equilibrium distribution, where the parameters $\eta$ and $T$ defining the Maxwellian distribution are dependent upon $z$, in our case (analogous to a global thermal equilibrium) the parameters of are independent of $z$, so we also integrate the kinetic equation over $z$ as well as the `momentum-like' variables (we note that we have already integrated the kinetic equation over $z$ when it is bounce-averaged). We also need only take moments over the trapped region of phase-space since the passing electron distribution is known.

As we seek an equation for $T$ it serves to take the $\mathcal{E}$-moment over trapped phase-space. The line-integrated energy density of trapped electrons is given by
\begin{equation}
    W_t = \int_{V_t} \mathcal{E} f\,\mathrm{d}^3v\,\mathrm{d}z, \label{eq:W_t}
\end{equation}
where $V_t$ is the trapped region of phase-space. Taking the $\mathcal{E}$-moment of Eq.\,\eqref{eq:kinetic_expansion} yields
\begin{equation}
        \frac{\mathrm{d} W_t}{\mathrm{d} t} = \frac{\mathrm{d}
        W_t}{\mathrm{d} t}\bigg|_\mathrm{adiabatic} + \frac{\mathrm{d}
        W_t}{\mathrm{d} t}\bigg|_\mathrm{separatrix} + \frac{\mathrm{d}
        W_t}{\mathrm{d} t}\bigg|_\mathrm{heating} \label{eq:dWtdt}
\end{equation}
for
\begin{align}
        &\frac{\mathrm{d} W_{t}}{\mathrm{d} t}\bigg|_\mathrm{adiabatic} &=
        -\frac{2\pi}{m_e^2}\int^\infty_0 \int^{0}_{-
        e\phi_m} \frac{\partial J}{\partial t}
        f\,\mathrm{d}\mathcal{E}_\parallel\,\mathrm{d}\mathcal{E}_\perp, \\
        &\frac{\mathrm{d} W_{t}}{\mathrm{d} t}\bigg|_\mathrm{separatrix}
        &= \frac{2\pi}{m_e^2}\frac{\partial J_m}{\partial
        t}\int^\infty_0 \mathcal{E}_\perp f_a\big|_{\mathcal{E}_\parallel = 0}\,\mathrm{d}\mathcal{E}_\perp, \\
        &\frac{\mathrm{d} W_{t}}{\mathrm{d} t}\bigg|_\mathrm{heating} &= \int_{V_t} \mathcal{E} C(f,f_\mathrm{II} + f_\mathrm{III})\,\mathrm{d}^3v\,\mathrm{d}z.
\end{align}
The details of the procedure are contained in Appendix A.

The terms in Eq.\,\eqref{eq:dWtdt} are descriptive: `adiabatic' corresponds to the adiabatic change in the electron energy as the well changes shape, `separatrix' corresponds to electrons crossing the trapped-passing separatrix due to the changing depth of the potential well $e\phi_m$, and `heating' corresponds to collisions of $f$ with the hot electrons.

Although $W_t$ is in principle expressible in terms of $T$, any tiny change in $W_t$ (and hence $T$) results in a huge deviation from quasineutrality due to the exponential dependency of $n_0$ on $e\phi/T$. Solving the evolution equation for $W_t$ numerically, inverting the relation to find $T$, and maintaining quasineutrality requires an impractically small timestep. Instead, we derive an energy conservation law for electrons and ions which can be used as an evolution equation for $T$.

The energy conservation law will contain contributions from both passing and trapped electrons, so it is more convenient to consider the energy contained within some interval $z \in [-L_S/2,L_S/2]$ for $L_S$ much larger than the plasmoid. The ambient plasma then acts as an infinite source and sink of electrons and energy for this section of the field line. The passing distribution function on this interval being $f_a$ can be understood as the ambient plasma instantly replenishing the passing distribution if it is altered in any way by interaction with the plasmoid; the ambient plasma essentially acts as a `thermostat' for the passing distribution. The interval $L_S$ must be large enough that the line-integrated trapped electron energy density $W_t$ changes negligibly as $L_S$ increases (i.e. the trapped electron density is negligible at $|z| = L_S/2$).

It is more convenient to work with the line-integrated \textit{kinetic} energy
\begin{equation}
    K_t = W_t + \int^\frac{L_S}{2}_{-\frac{L_S}{2}} e \phi n_{e,t}\,\mathrm{d}z
\end{equation}
(where $n_{e,t}$ is the trapped electron density) rather than $W_t$, since we expect the sum of kinetic energies to exhibit a conservation law. Taking the time derivative of $K_t$ yields
\begin{equation}
\begin{split}
    \frac{\mathrm{d} K_t}{\mathrm{d} t} &= \frac{\mathrm{d} W_t}{\mathrm{d}t} + \int^\frac{L_S}{2}_{-\frac{L_S}{2}} e \frac{\partial \phi}{\partial t}n_{e,t}\,\mathrm{d}z + \int^\frac{L_S}{2}_{-\frac{L_S}{2}} e\phi \frac{\partial n_{e,t}}{\partial t}\,\mathrm{d}z  \\
    &= \frac{\mathrm{d} W_t}{\mathrm{d}t} - \frac{\mathrm{d} W_t}{\mathrm{d} t}\bigg|_\mathrm{adiabatic} + \int^\frac{L_S}{2}_{-\frac{L_S}{2}} e\phi \left(Z\frac{\partial n_i}{\partial t} - \frac{\partial n_{e,p}}{\partial t}\right)\,\mathrm{d}z,
\end{split}
\end{equation}
where we have used the quasineutrality condition (assuming that there is a single species of ions with charge $Z$) and the fact that $n_{e,t} + n_{e,p} = n_e$ for $n_{e,p}$ the passing electron density. The latter two terms correspond to the changing kinetic energies of the ions and passing electrons. The term involving ion density is given by
\begin{equation}
 \int^\frac{L_S}{2}_{-\frac{L_S}{2}} Z e \phi \frac{\partial n_i}{\partial t}\,\mathrm{d}z = -\frac{\partial K_i}{\partial t}
\end{equation}
for $K_i$ the line-integrated kinetic energy of the ions. This can be derived by taking velocity moments Eq.\,\eqref{eq:ion_vlasov} or by constructing an energy equation from Eqs.\,\eqref{eq:ion_continuity},\eqref{eq:ion_velocity}. The same procedure cannot be carried out for the passing electrons since these are restricted to a certain region of phase space. 

Extending the notion of an orbit integral to electrons with positive parallel energy is straightforward on a finite $z$ interval:
\begin{equation}
    \oint g(\mathcal{E}_\parallel > 0)\,\mathrm{d}z = 2 \int^\frac{L_S}{2}_{-\frac{L_S}{2}} g(\mathcal{E}_\parallel > 0)\,\mathrm{d}z,
\end{equation}
allowing us to define the second adiabatic invariant $J$ for positive parallel energies. The line-integrated passing electron energy density on the interval $z \in [-L_S/2,L_S/2]$ is given by
\begin{equation}
      W_p = \frac{2\pi}{m_e^2}\int^\infty_0\int^\infty_{J_m} \mathcal{E} f_a\,\mathrm{d}J\,\mathrm{d}\mathcal{E}_\perp,
\end{equation}
hence
\begin{equation}
    \frac{\mathrm{d} W_p}{\mathrm{d} t} = \frac{2\pi}{m_e^2}\int^\infty_0 \int^\infty_0 \left(\frac{\mathcal{E}}{T_a} -  1\right)\frac{\partial J}{\partial t} f_a\,\mathrm{d}\mathcal{E}_\parallel\,\mathrm{d}\mathcal{E}_\perp
    - \frac{2\pi}{m_e^2}\int^\infty_0 \mathcal{E}_\perp \frac{\mathrm{d} J_m}{\mathrm{d} t}f_a(\mathcal{E}_\perp) \,\mathrm{d}\mathcal{E}_\perp.
\end{equation}
We note that
\begin{equation}
    \frac{2\pi}{m_e^2}\int^\infty_0 \int^\infty_0 \frac{\partial J}{\partial t} f_a\,\mathrm{d}\mathcal{E}_\parallel\,\mathrm{d}\mathcal{E}_\perp = - \int^\frac{L_S}{2}_{-\frac{L_S}{2}} e \frac{\partial \phi}{\partial t}n_{e,p}\,\mathrm{d}z
\end{equation}
and
\begin{equation}
    \frac{2\pi}{m_e^2}\int^\infty_0 \mathcal{E}_\perp \frac{\mathrm{d} J_m}{\mathrm{d} t}f_a(\mathcal{E}_\perp) \,\mathrm{d}\mathcal{E}_\perp = \frac{\mathrm{d} W_t}{\mathrm{d} t}\bigg|_\mathrm{separatrix},
\end{equation}
so, given that 
\begin{equation}
    K_p = W_p + \int^\frac{L_S}{2}_{-\frac{L_S}{2}} e\phi n_{e,p}\,\mathrm{d}z,
\end{equation}
we obtain the energy conservation law on the interval $z \in [-L_S/2,L_S/2]$
\begin{equation}
    \frac{\mathrm{d}}{\mathrm{d} t}\left(K_t + K_p + K_i\right) = \frac{\mathrm{d} W_t}{\mathrm{d} t}\bigg|_\mathrm{heating} + \frac{2\pi}{m_e^2}\int^\infty_0 \int^\infty_0 \frac{\mathcal{E}}{T_a}\frac{\partial J}{\partial t} f_a\,\mathrm{d}\mathcal{E}_\parallel\,\mathrm{d}\mathcal{E}_\perp.
\end{equation}
Evidently, the inclusion of the ion and passing electron energies in the energy conservation law accounts for the absence of $\mathrm{d} W_t/\mathrm{d}t\big|_\mathrm{adiabatic}$, since this represents energy that is extracted from trapped electrons during the expansion and given to other species. The absence of the separatrix term is simply due to this contribution cancelling out between $K_t$ and $K_p$.

The second term on the right hand side of the above arises from the fact that passing electrons have their energy altered when passing through the time-varying potential well, and this energy gain (or loss) is absorbed by (or suffered by) the ambient plasma, which continually supplies passing electrons following the distribution $f_a$. The heating due to this effect is essentially negligible when $T \ll T_a$, but provides a considerable fraction of the heating power when $T \sim T_a$.

The first term on the right hand side represents the collisional heating of trapped electrons. In \cite{Arnold2023}, which treated a high-$Z$ plasmoid, it was shown that the heating rate for cold electrons in a plasmoid was 3/4 that expected for cold electrons in a homogeneous plasma, since the acceleration of passing electrons through the potential well decreases their density and collisionality. That is, given that the per-electron heating rate of a cold Maxwellian in a homogeneous plasma is $3\nu_h(T_a-T)$, the per-electron heating rate for electrons trapped in the potential well was found to be $(9/4) \nu_h(T_a-T)$. In our case the distribution function is also highly isotropic for $\mathcal{E} < 0$, so we approximate the collisional heating term by
\begin{equation}
    \frac{\mathrm{d} W_t}{\mathrm{d} t}\bigg|_\mathrm{heating}
    = \frac{9}{4} \nu_h N_{\mathcal{E}<0}(T_a - T),
\end{equation}
where $N_{\mathcal{E}<0}$ is the line-integrated density of trapped electrons with $\mathcal{E} < 0$. As shown in Section \ref{sec:qe_numerical}, the distribution function is somewhat less than $f_a$ in the $\mathcal{E} > 0$ region for $Z = 1$, so the above is an upper bound for the heating rate.

Consolidating trapped and passing electron energies into $K_e = K_t + K_p$ we have the energy conservation law
\begin{equation}
    \frac{\mathrm{d}}{\mathrm{d}t}\left(K_e + K_i\right) = Q \label{eq:energy_balance}
\end{equation}
for heating power
\begin{equation}
    Q = \frac{9}{4}\nu_hN_{\mathcal{E}<0}(T_a - T) + \frac{2\pi}{m_e^2}\int^\infty_0 \int^\infty_0 \frac{\mathcal{E}}{T_a}\frac{\partial J}{\partial t} f_a\,\mathrm{d}\mathcal{E}_\parallel\,\mathrm{d}\mathcal{E}_\perp.
    \label{eq:heating}
\end{equation}

\subsection{Comparison of the system with earlier work}
At this point a clear comparison can be drawn between this investigation and \cite{Aleynikov2019,Arnold2021}. In \cite{Aleynikov2019,Arnold2021} a cold-fluid system for ions was coupled with an energy conservation law for the plasmoid ions and electrons. The dynamics of the passing electron distribution were neglected save for the assertion that the presence of the ambient plasma resulted in a per-electron heating term $3\nu_h T_a$ for plasmoid electrons (only the linear heating stage was treated). The potential was given by the Boltzmann relation and was unbounded as $|z| \rightarrow \infty$ due to the neglect of passing electrons.

Here, we use an energy conservation law with modified heating terms and electron kinetic energy derived from a solution to the electron kinetic problem. The electric potential is given by the quasineutrality condition according to the electron density of this distribution, and vanishes as $|z| \rightarrow \infty$. Taking the limit $n_a \rightarrow 0$ in Eq.\,\eqref{eq:QN} and the energy conservation law \eqref{eq:energy_balance} recovers the same system as \cite{Aleynikov2019} with the exception of the 3/4 factor on the collisional heating term.

\subsection{Numerical solutions of the plasmoid expansion system} \label{sec:expansion_numerical}
Numerical solutions to the system created by coupling the energy conservation law \eqref{eq:energy_balance}, quasineutrality \eqref{eq:QN_2}, and one of systems describing the ions (the Vlasov equation \eqref{eq:ion_vlasov} or the cold-ion system \eqref{eq:ion_continuity},\eqref{eq:ion_velocity}) were obtained using the plasma parameters $n_a = 5\times10^{19}\,\mathrm{m}^{-3}$, $T_a = 5\,\mathrm{keV}$ and $N_{ic} = 10^{22}\,\mathrm{m}^{-2}$, the same as in Section \ref{sec:qe_numerical}. The plasma was once more assumed to be hydrogenic. The heating timescale with these parameters is $\nu_h^{-1} = 162\,\mathrm{\mu s}$ and the expansions were run to $300\,\mathrm{\mu s}$, at which point the plasmoid and ambient temperatures have nearly equalised and the plasmoid density has dropped to nearly the ambient. 

In the collisionless ion expansion the ambient ion distribution was initialised to a Maxwellian of density $n_a$ and temperature $T_a$. The plasmoid ions were initialised at a temperature of $50\,\mathrm{eV}$. In both the cold-fluid and collisionless expansion the plasmoid was initialised in the lowest-$z$ grid cell and the ambient uniformly across $z$

The analytical solution to the QE problem given in Section \ref{sec:qe_approx} was used to calculate densities and energy densities, respectively appearing in the quasineutrality condition \eqref{eq:QN_2} and the energy conservation law \eqref{eq:energy_balance}. We neglect any deviation of $f$ from $f_0$ in the $\mathcal{E}<0$ region when evaluating these densities as the difference is negligible. However, we do take into account the fact that $f$ is flattened in the $\mathcal{E} > 0$, $\mathcal{E}_\parallel < 0$ region, as this will have a relatively large impact on the density and energy density. We use Eq.\,\eqref{eq:eta_approx} as the expression for $\eta$ in terms of $T$ when evaluating $f_0$. Additionally, we expand the $\mathrm{cosh}$ functions found in the analytical expression of $f$ to second order in order to obtain analytical expressions for the moments.

The electron density is then given by
\begin{equation}
    n_e = n_{\mathcal{E}<0} +
    n^{\mathcal{E}>0,c}_{\mathcal{E}_\parallel < 0} + n^{\mathcal{E}>0,a}_{\mathcal{E}_\parallel < 0} + 
    n_{\mathcal{E}_\parallel > 0} \label{eq:ne_approx}
\end{equation}
for
\begin{equation}
    n_{\mathcal{E}<0} = \eta \left(\mathrm{e}^\frac{e\phi}{T}\mathrm{erf}\left(\sqrt{\frac{e\phi}{T}}\right) - \frac{2}{\sqrt{\pi}}\sqrt{\frac{e\phi}{T}}\right),\label{eq:ne_energytrapped}
\end{equation}
\begin{equation}
    n^{\mathcal{E}>0,c}_{\mathcal{E}_\parallel < 0} = \frac{2}{\sqrt{\pi}}\eta\left[\sqrt{\frac{e\phi}{T}} - \frac{1}{1 + \frac{2}{1+Z}\frac{e\phi_m}{T}}\left(\sqrt{\frac{e\phi}{T}} + \frac{1}{3}\frac{2}{1+Z}\left(\frac{e\phi}{T}\right)^\frac{3}{2}\right)\right],
\end{equation}
\begin{equation}
   n^{\mathcal{E}>0,a}_{\mathcal{E}_\parallel < 0} = \frac{2}{\sqrt{\pi}}n_a \frac{1}{1 + \frac{2}{1+Z}\frac{e\phi_m}{T_a}}\left(\sqrt{\frac{e\phi}{T_a}} + \frac{1}{3}\frac{2}{1+Z}\left(\frac{e\phi}{T_a}\right)^\frac{3}{2}\right),
\end{equation}
and
\begin{equation}
    n_{\mathcal{E}_\parallel > 0} = n_a \mathrm{e}^\frac{e\phi}{T_a}\mathrm{erfc}\left(\sqrt{\frac{e\phi}{T_a}}\right).
\end{equation}

The line-integrated kinetic energies are given by
\begin{equation}
    K_e = K_{\mathcal{E} < 0} +
    K^{\mathcal{E}>0,c}_{\mathcal{E}_\parallel < 0} + K^{\mathcal{E}>0,a}_{\mathcal{E}_\parallel < 0} +
    K_{\mathcal{E}_\parallel > 0}, \label{eq:Ke_approx}
\end{equation}
where
\begin{equation}
    K_{\mathcal{E}<0} = \frac{3}{2}N_{\mathcal{E}<0}T -\frac{2}{\sqrt{\pi}}\eta\int^\frac{L_S}{2}_{-\frac{L_S}{2}}\left(\frac{e\phi}{T}\right)^\frac{3}{2}\,\mathrm{d}z,
\end{equation}
\begin{equation}
    K^{\mathcal{E}>0,c}_{\mathcal{E}_\parallel < 0} = N^{\mathcal{E}>0,c}_{\mathcal{E}_\parallel < 0}T + \int^\frac{L_S}{2}_{-\frac{L_S}{2}}e\phi n^{\mathcal{E}>0,c}_{\mathcal{E}_\parallel < 0}\,\mathrm{d}z,
\end{equation}
\begin{equation}
    K^{\mathcal{E}>0,a}_{\mathcal{E}_\parallel < 0} = N^{\mathcal{E}>0,a}_{\mathcal{E}_\parallel < 0}T_a + \int^\frac{L_S}{2}_{-\frac{L_S}{2}}e\phi n^{\mathcal{E}>0,a}_{\mathcal{E}_\parallel < 0}\,\mathrm{d}z,
\end{equation}
and
\begin{equation}
    K_{\mathcal{E}_\parallel > 0} = \frac{3}{2}N_{\mathcal{E}_\parallel > 0}T_a + \frac{1}{\sqrt{\pi}}n_a T_a \int^\frac{L_S}{2}_{-\frac{L_S}{2}}\sqrt{\frac{e\phi}{T_a}}\,\mathrm{d}z,
\end{equation}
where
\begin{align}
    N_{\mathcal{E}<0} = &\int^\frac{L_S}{2}_{-\frac{L_S}{2}}n_{\mathcal{E}<0}\,\mathrm{d}z, \\
    N^{\mathcal{E}>0,c}_{\mathcal{E}_\parallel < 0} = &\int^\frac{L_S}{2}_{-\frac{L_S}{2}}n^{\mathcal{E}>0,c}_{\mathcal{E}_\parallel < 0}\,\mathrm{d}z, \\
    N^{\mathcal{E}>0,a}_{\mathcal{E}_\parallel < 0} = &\int^\frac{L_S}{2}_{-\frac{L_S}{2}}n^{\mathcal{E}>0,a}_{\mathcal{E}_\parallel < 0}\,\mathrm{d}z, \\
    N_{\mathcal{E}_\parallel > 0} = &\int^\frac{L_S}{2}_{-\frac{L_S}{2}}n_{\mathcal{E}_\parallel > 0}\,\mathrm{d}z.
\end{align}

\begin{figure}
    \centering
    \hspace{0.5cm}
    \includegraphics[width=\textwidth]{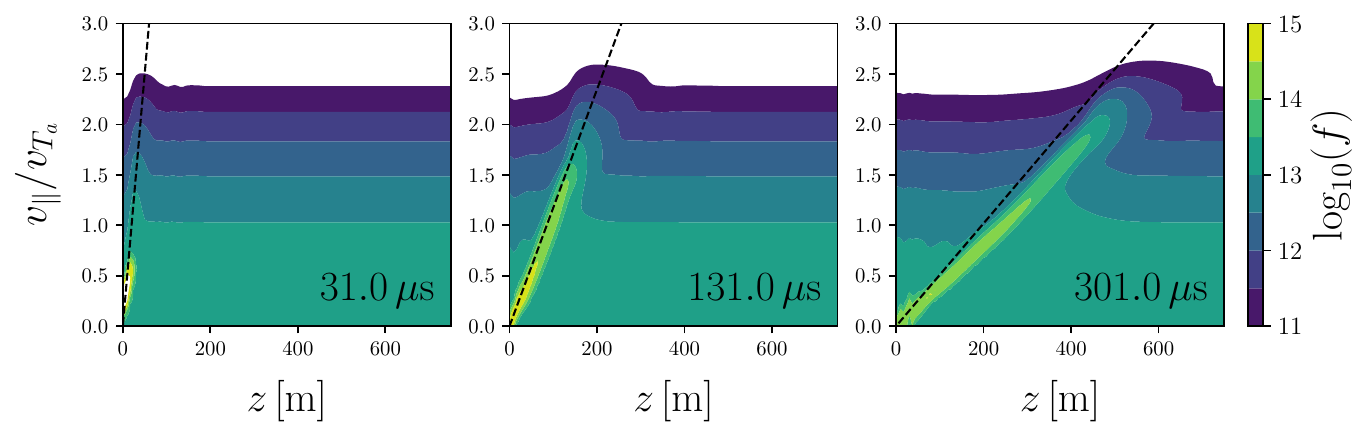}
    \caption{Ion distribution function of the collisionless ion expansion at various times. The dashed line is the self-similar flow velocity given in \cite{Aleynikov2019}.}
    \label{fig:1e22-kinetic-hota-fi}
\end{figure}

\begin{figure}
    \centering
    \hspace{0.5cm}
    \includegraphics[width=\textwidth]{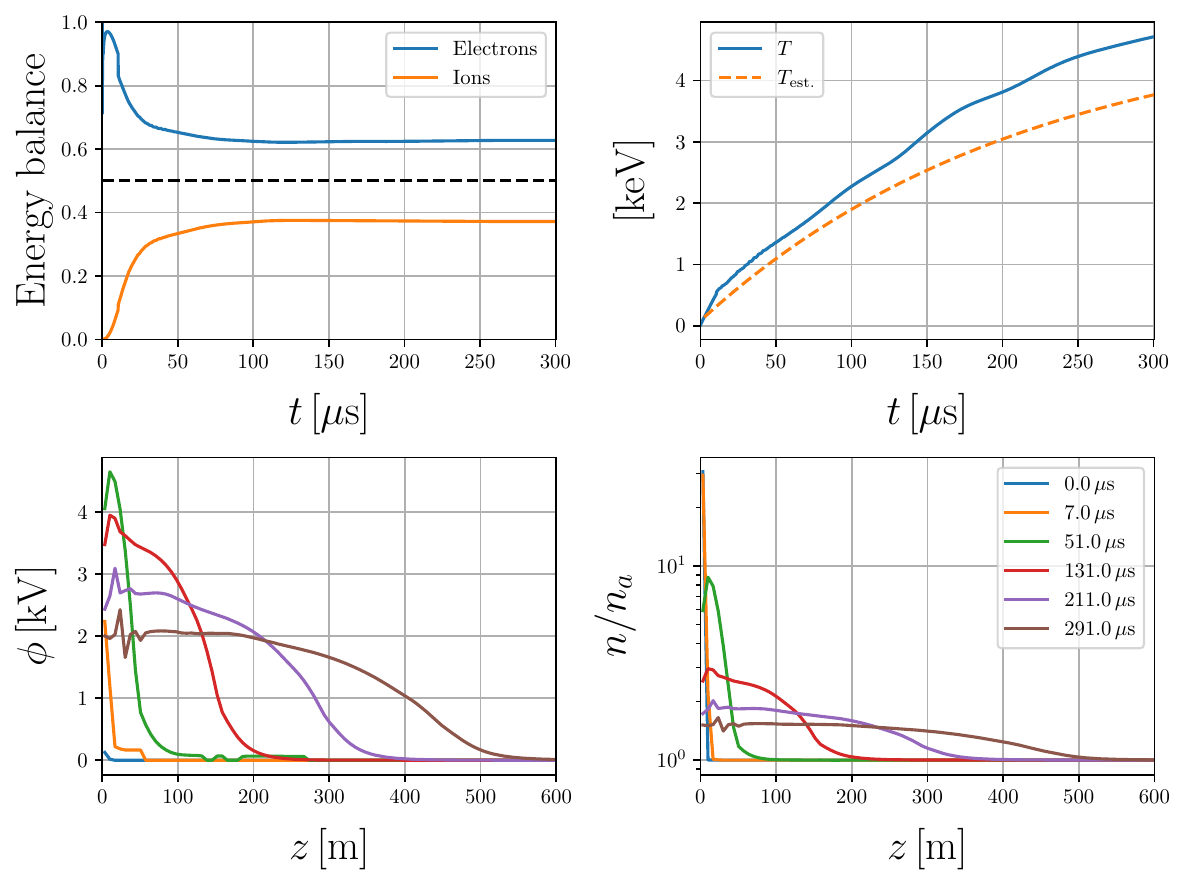}
    \caption{Derived quantities of the collisionless ion expansion. Top left: the relative amounts of energy deposited into the electrons and ions. Top right: the plasmoid electron temperature $T$ and the estimated temperature evolution given a homogeneous plasma. Bottom: plots of the electric potential and electron density at various times.}
    \label{fig:1e22-kinetic-hota-derived}
\end{figure}

\begin{figure}
    \centering
    \hspace{0.5cm}
    \includegraphics[width=\textwidth]{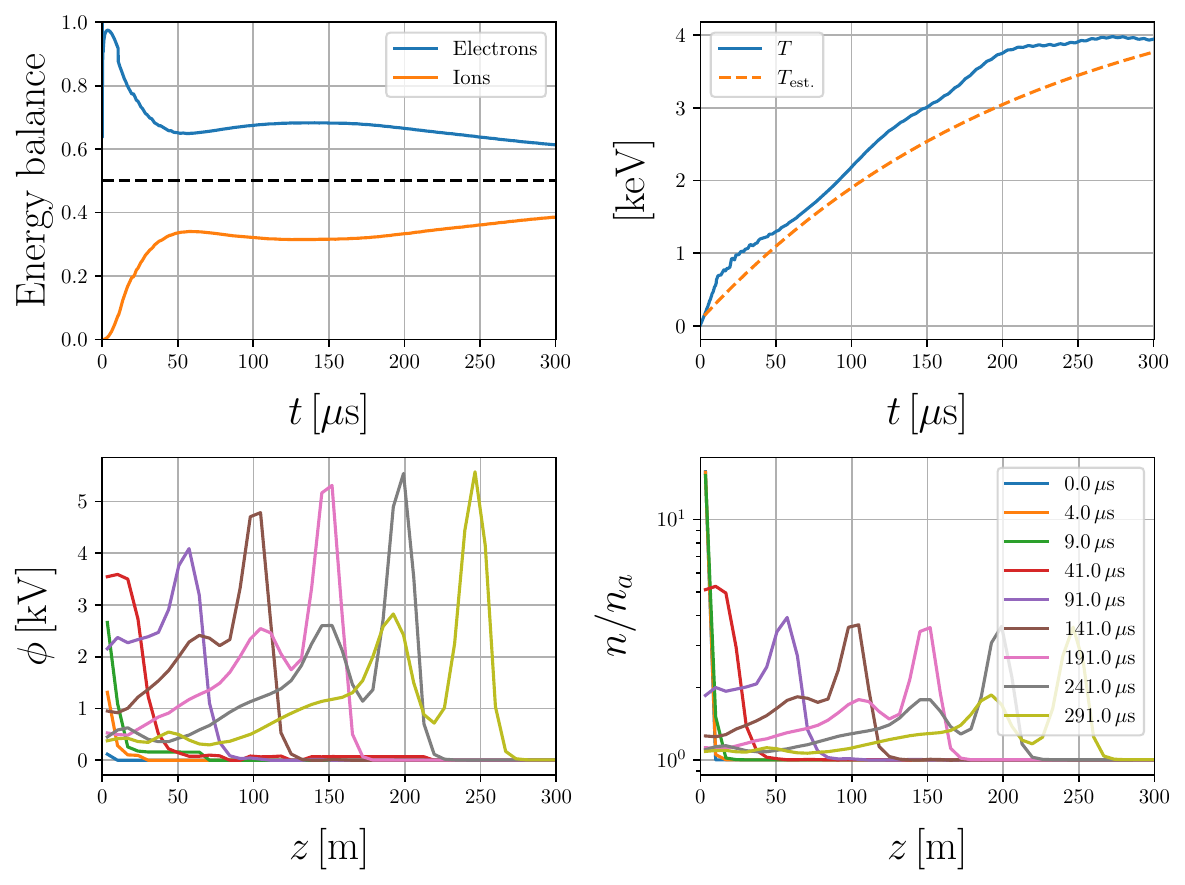}
    \caption{Derived quantities of the cold-fluid ion expansion. Top left: the relative amounts of energy deposited into the electrons and ions. Top right: the plasmoid electron temperature $T$ and the estimated temperature evolution given a homogeneous plasma. Bottom: plots of the electric potential and electron density at various times.}
    \label{fig:1e22-fluid-derived}
\end{figure}

Figure \ref{fig:1e22-kinetic-hota-fi} shows the ion distribution of the collisionless ion expansion at various times. We see that the self-similar flow velocity matches in the regions of highest ion density even at late times. Figures \ref{fig:1e22-kinetic-hota-derived} and \ref{fig:1e22-fluid-derived} show quantities derived from the ion distribution or from the flow velocities and densities of the cold-fluid expansion. In the fluid expansion an ion front rapidly forms, resulting in very large density gradients. The front appears to develop an oscillatory character at later times. Sound waves and perhaps even solitons would develop near the ion front and propagate into the ambient plasma, but the $z$ grid used was too course to resolve such waves. In contrast the collisionless expansion exhibits no steep front owing to the fact that the hot ambient plasma does not `pile-up' on the moving plasmoid; the majority of passing electrons have large enough parallel velocity to either pass over the plasmoid or be reflected from the front much more quickly than the expansion. In both cases the electron temperature approaches $T_a$ somewhat more quickly than the estimate \eqref{eq:T_Aleynikov_modified}, a consequence of the term in Eq.\,\eqref{eq:heating} corresponding to the energy lost adiabatically by the passing electrons but gained by trapped electrons and ions. This term constitutes the majority of the heating when $T \sim T_a$.

Of particular note, and ultimately what is sought in this investigation, is the energy balance between electrons and ions, plotted in the top left of Figs.\,\ref{fig:1e22-kinetic-hota-derived} and \ref{fig:1e22-fluid-derived}. The two lines represent the fraction of the heating energy ($\int^t_0Q\,\mathrm{d}t$ for $Q$ in Eq.\,\eqref{eq:energy_balance}) deposited into each species, and in all cases the energy balance tends to a near equal split between electrons and ions, which is quite remarkable considering the opposite collisionality regimes for ions in the fluid and Vlasov models. This energy balance is similar to those predicted by the self-similar expansions in \cite{Aleynikov2019} and \cite{Arnold2021}.

Although not shown here, numerical simulations of the expansion with larger line-integrated plasmoid densities exhibit an energy balance even closer to an equal split. A larger line-integrated plasmoid density also results in the plasmoid and ambient temperature equilibrating before the plasmoid density drops to the ambient, which furnishes the possibility of modelling the expansion (and wave generation) from that point with the much simplified equations resulting from a Maxwellian electron distribution function.

\FloatBarrier

\section{Discussion and Conclusions}
A model for the expansion of a plasmoid produced by fuel pellet ablation has been developed that, considering the high complexity of the problem, provides a relatively simple steady-state electron kinetic problem that was rigorously derived from the ordering of timescales. The long-term evolution of the electrons is decsribed by an energy balance obtained by taking moments of the electron kinetic equation. Ions were described with cold-fluid equations or with the Vlasov equation; in the latter case the large temperature difference between ambient and plasmoid ions can be accounted for.

[Discussion of shock; happens when $T_a$ is small]

The model presented in this paper is contrasted with earlier work that simply assumed a Maxwellian electron distribution for plasmoid electrons and entirely neglected ambient electrons except in a simple heating term. Earlier work also only considered a cold-fluid model for the ions. The largest pitfall of earlier work was the combination of an unbounded (and therefore unphysical) electric potential and the lack of a proper treatment of passing electrons, which called into question the conclusions about the qualitative character of the expansion, most notably the electron to ion energy transfer. The rectification of these issues was the main motivation for developing a model in a more rigorous manner.

It has been shown that during the expansion electrons reach a `quasi-equilibrium' state: a dynamical steady-state on the fastest collisional timescale which establishes an electron distribution that has no net flux through the trapped-passing separatrix. An analytical solution to the QE electron kinetic problem was obtained and compared to a numerical solution. An estimate of the height of the self-consistent electric potential that supports quasi-equilibrium has been derived. The estimate is consistent with the Boltzmann relation when the temperatures of the plasmoid electrons and ambient electrons are equal, and is in agreement with the height of the self-consistent potential found in the solution to the time-dependent electron kinetic problem in a high-$Z$ plasmoid \citep{Arnold2023}, providing strong evidence for the establishment of the QE state. The QE kinetic equation and energy balance can therefore be incorporated into established codes to describe electron dynamics in a pellet plasmoid.

The quasineutrality condition, no-net-flux condition, and QE kinetic problem allow a description of the QE distribution function and electric potential in terms of the plasmoid electron temperature $T$ and the ion densities. Analogous to the Braginskii equations, the evolution equation for $T$ was obtained by taking the energy moment over the electron kinetic equation that holds on the expansion timescale; this evolution equation takes the form of an energy conservation law for both electrons and ions. The evolution of $T$ is driven by the energy exchange between passing electrons, trapped electrons, and ions; heating power initially deposited in the plasmoid electrons by collisions with the hot ambient electrons can be redistributed between the species.

When modelling the expansion collisionless and fluid models for ions were used because of their opposite collisionality regimes; it is expected that the shared qualitative properties of these expansions hold with a more accurate model for the ions. The most important qualitative feature of the expansions is the near-equal split of the heating power between electrons and ions. This energy balance is in agreement with that of the self-similar solutions to the expansion found in \cite{Aleynikov2019,Arnold2021}. The explanation for this result is that self-similar solutions tend to be `attractors' for more complicated systems, with the particularly robust predictions being those that do not contain reference to any parameters at all, such as the energy balance \citep{Barenblatt1996}. It is reasonable therefore to suggest that the energy balance holds with a more accurate model of the ions, perhaps also one including motion transverse to magnetic field lines, which would allow a description of the assimilation of the plasmoid into the core of the device.

Since this energy balance entails a considerable transfer of energy from electrons to ions we conclude that the ambipolar expansion of a pellet plasmoid is a potent mechanism for the heating of ions on a much faster timescale than that on which electron-ion collisions occur; the expansion happens on the hot electron-hot electron collision timescale and the resulting ion flow energy is converted to thermal energy on the ion-ion collision timescale, which is approximately $\sqrt{m_i/m_e}$ times smaller than the electron-ion collision timescale. Hence fuel pellet injection should be considered as not just a method for replenishing lost plasma, but also as a technique for rapidly heating ions if their temperature is exceeded by that of the electrons.

\appendix
\numberwithin{equation}{section}
\section{Calculation of the $\mathcal{E}$-moment of the electron kinetic equation on the expansion timescale}
The volume element in the variables $(\mathcal{E}_\parallel, \mathcal{E}_\perp, z)$ is given by
\begin{equation}
        \mathrm{d}^3v\,\mathrm{d}z =
        \frac{4\pi}{m_e^2}\frac{1}{v_\parallel}\,\mathrm{d}\mathcal{E}_\parallel\,\mathrm{d}\mathcal{E}_\perp\,\mathrm{d}z,
\end{equation}
so phase-space integrals over the trapped domain $V_t$ may be expressed as
\begin{equation}
        \begin{split}
        \int_{V_t} h\,\mathrm{d}v^3\,\mathrm{d}z &= \frac{4\pi
        }{m_e^2}\int^\infty_{-\infty}\int^\infty_0\int^{
        0}_{-e\phi}\frac{h}{v_\parallel}
\,\mathrm{d}\mathcal{E}_\parallel\,\mathrm{d}\mathcal{E}_\perp\,\mathrm{d}z \\
        &= \frac{2\pi}{m_e^2}\int^\infty_0\int^{0}_{-e\phi_m} \oint
                \frac{h}{v_\parallel}\,\mathrm{d}z\,\mathrm{d}\mathcal{E}_\parallel\,\mathrm{d}\mathcal{E}_\perp
        \\
        &= \frac{2\pi}{m_e^2}\int^\infty_0\int^{0}_{-e\phi_m}
                \left<h\right>\tau
                \,\mathrm{d}\mathcal{E}_\parallel\,\mathrm{d}\mathcal{E}_\perp, \\
        &= \frac{2\pi}{m_e^2}\int^\infty_0\int^{J_m}_{0}
                \left<h\right>
                \,\mathrm{d}J\,\mathrm{d}\mathcal{E}_\perp.
        \end{split} \label{eq:intermediate_2}
\end{equation}
Hence, for a term $g$ that is already bounce-averaged, the integral over phase-space is given by
\begin{equation}
    \frac{2\pi}{m_e^2} \int^\infty_0 \int^0_{-e\phi_m} \tau g\,\mathrm{d}\mathcal{E}_\parallel\,\mathcal{E}_\perp = \frac{2\pi}{m_e^2}\int^\infty_0 \int^{J_m}_0 g\,\mathrm{d}J\mathrm{d}\mathcal{E}_\perp.
\end{equation}

We note that Eq.\,\eqref{eq:kinetic_expansion} may be expressed as
\begin{equation}
    \frac{\partial f}{\partial t}\bigg|_J = \left<C(f,f_\mathrm{II} + f_\mathrm{III})\right>,
\end{equation}
where $\cdot|_J$ indicates a derivative at constant $J$ rather than constant $\mathcal{E}_\parallel$. Hence its integral over phase-space (after being multiplied by $\mathcal{E}$) may be expressed as
\begin{equation}
        \frac{2\pi}{m_e^2}\int^\infty_0 \int^{J_m}_0\mathcal{E} \frac{\partial
        f}{\partial t}\bigg|_J\,\mathrm{d}J\,\mathrm{d}\mathcal{E}_\perp = \int_{V_t} \mathcal{E}C(f,f_\mathrm{II} + f_\mathrm{III})\,\mathrm{d}^3v\,\mathrm{d}z,
        \label{eq:intermediate_1}
\end{equation}
where we have used the fact
\begin{equation}
    \frac{2\pi}{m_e^2}\int^\infty_0 \int^{J_m}_0 \mathcal{E}\left<C(f,f_\mathrm{II} + f_\mathrm{III})\right>\,\mathrm{d}J\,\mathrm{d}\mathcal{E}_\perp = \int_{V_t} \mathcal{E}C(f,f_\mathrm{II} + f_\mathrm{III})\,\mathrm{d}^3v\,\mathrm{d}z.
\end{equation}

From Eq.\,\eqref{eq:W_t} we find
\begin{equation}
        \begin{split}
        \frac{\mathrm{d} W_t}{\mathrm{d} t} =
        -\frac{2\pi}{m_e^2}&\int^\infty_0\int^{0}_{-e\phi_m}\frac{\partial
        J}{\partial t}
f\,\mathrm{d}\mathcal{E}_\parallel\,\mathrm{d}\mathcal{E}_\perp
        + \frac{2\pi}{m_e^2}\frac{\partial
        J_m}{\partial t}\int^\infty_0 \mathcal{E}_\perp
        f_a\big|_{\mathcal{E}_\parallel = 0} \,\mathrm{d}\mathcal{E}_\perp \\&+
        \int_{V_t} \mathcal{E}C(f,f_\mathrm{II} + f_\mathrm{III})\,\mathrm{d}^3v\,\mathrm{d}z,
        \end{split}
\end{equation}
where we have used the fact that
\begin{equation}
    \frac{\partial \mathcal{E}}{\partial t}\bigg|_J = -\frac{1}{\tau}\frac{\partial J}{\partial t}\bigg|_{\mathcal{E}_\parallel}.
\end{equation}

\section*{Funding}
This work has been carried out within the framework of the EUROfusion Consortium, funded by the European Union via the Euratom Research and Training Programme (Grant Agreement No 101052200 -- EUROfusion). Views and opinions expressed are however those of the author(s) only and do not necessarily reflect those of the European Union or the European Commission. Neither the European Union nor the European Commission and be held responsible for them. This work was supported by the U.S. Department of Energy under Contract Nos. DEFG02-04ER54742 and DESC0016283.

\section*{Declaration of interests}
The authors report no conflict of interest.

\bibliography{master}
\bibliographystyle{authordate1}
\end{document}